\documentclass[onecolumn,twoside]{IEEEtran}
\usepackage{mathpazo}
\usepackage{times}
\usepackage{amsmath}
\usepackage{amsfonts}
\usepackage{latexsym}
\usepackage{amssymb}
\usepackage{mathrsfs}
\usepackage{upref}
\usepackage{theorem}
\usepackage{graphicx}
\usepackage{psfrag}
\usepackage{cite}
\usepackage{multirow}
\usepackage{color}

\theoremstyle{plain}
\theorembodyfont{\normalfont\slshape}

\newtheorem{thm}{Theorem$\!$}
\newenvironment{theorem}
{\begin{thm}\hspace*{-1ex}{\bf.}}{\end{thm}}

\newtheorem{lem}[thm]{Lemma$\!$}
\newenvironment{lemma}{\begin{lem}\hspace*{-1ex}{\bf.}}{\end{lem}}

\newtheorem{alg}[thm]{Algorithm$\!$}

\newtheorem{prop}[thm]{Proposition$\!$}

\newtheorem{cor}[thm]{Corollary$\!$}
\newenvironment{corollary}{\begin{cor}\hspace*{-1ex}{\bf.}}{\end{cor}}

\newtheorem{defn}[thm]{Definition$\!$}
\newenvironment{definition}{\begin{defn}\hspace*{-1ex}{\bf.}}{\end{defn}}

\newtheorem{xmpl}[thm]{Example$\!$}
\newenvironment{example}{\begin{xmpl}\hspace*{-1ex}{\bf.}}{\hfill$\Box$\end{xmpl}}

\newtheorem{cnstr}{Construction$\!$}

\newcounter{enumrom}
\renewcommand{\theenumrom}{(\roman{enumrom})}

\makeatletter
\renewcommand{\@endtheorem}{\endtrivlist}
\makeatother

\makeatletter
\renewcommand{\thefigure}{{\@arabic\c@figure}}
\renewcommand{\fnum@figure}{{\bf Figure\,\thefigure}}
\makeatother

\newcommand{\cH}{\mathcal{H}}

\newcommand{\cO}{\mathcal{O}}

\newcommand{\pf}{{\bf Proof: }}

\newcommand{\uw}{\mbox{$\underline{w}$}}
\newcommand{\uv}{\mbox{$\underline{v}$}}

\newcommand{\be}[1]{\begin{equation}\label{#1}}
\newcommand{\ee}{\end{equation}}

\renewcommand{\leq}{\leqslant}

\renewcommand{\geq}{\geqslant}

\newcommand{\Cref}[1]{Co\-ro\-lla\-ry\,\ref{#1}}

\newcommand{\C}{\mbox{${\cal C}$}}

\newcommand{\qed}{\hfill$\Box$\\[1ex]}
\newcommand{\hs}{\mbox{$\hat{s}$}}
\newcommand{\uc}{\mbox{$\underline{c}$}}
\newcommand{\uu}{\mbox{$\underline{u}$}}
\newcommand{\al}{\alpha}
\newcommand{\uzero}{\underline{0}}
\newcommand{\xor}{\oplus}

\newcommand{\eq}{\mbox{$\,=\,$}}
\newcommand{\ga}{\mbox{$\gamma$}}

\outer\def\proclaim #1. #2\par{\medbreak
 \noindent{\bf#1.\enspace}{\sl#2\par}%
 \ifdim\lastskip<\medskipamount \removelastskip\penalty55\medskip\fi}

\begin{document}


\title{\LARGE\bf Multiple-Layer Integrated Interleaved Codes: A Class
of Hierarchical Locally Recoverable Codes}

\author{\large
Mario~Blaum \\
IBM Research Division\\
Almaden Research Center\\
 San Jose, CA 95120, USA \\
Mario.Blaum@ibm.com}


\maketitle

\begin{abstract}
The traditional definition of Integrated Interleaved (II) codes
generally assumes that the component nested codes are either
Reed-Solomon (RS) or shortened Reed-Solomon codes. By
taking general classes of codes, we present a recursive construction
of Extended Integrated Interleaved (EII) codes into multiple layers,
a problem that brought attention in literature for II codes.
The multiple layer approach allows for a hierarchical scheme where
each layer of the code provides for a different locality. In
particular, we present the erasure-correcting capability of the new
codes and we show
that they are ideally suited as Locally Recoverable codes
(LRC) due to their hierarchical locality and the small finite field
required by the construction. Properties of the multiple layer EII
codes, like their minimum distance and dimension, as well as their
erasure decoding algorithms, parity-check matrices and performance
analysis, are provided and illustrated with examples. Finally, we will observe
that the parity-check matrices of high layer EII codes have low
density.
\end{abstract}


\section{Introduction}
\label{int}
The construction of $t$-level Integrated Interleaved
(II)~\cite{bh,hapkt,lxx,tk,w,z3} and
Extended Integrated Interleaved (EII)~\cite{b,bh2}
uses $t$ nested codes $\C_i$ over a finite field
$GF(q)$ (for simplicity, in this paper we assume that the field
$GF(q)$ has characteristic 2, but the constructions are valid over
fields of any characteristic). The idea is to divide $mn$ symbols
into $m$ distinct codewords, each codeword having a certain
correcting capability so they can be corrected locally. In addition,
the $m$ codewords share parity symbols that
enhance the correction capability of the individual codewords. There is a vast
literature on codes with such characteristics (see for
example~\cite{b} and references within). In particular, II and EII
codes are connected to Generalized Concatenated
codes~\cite{bz,z} and to Tensor Product codes~\cite{hys,hyus,if,wolf}.
In~\cite{bh}, $t$-level II codes were proposed as Locally Recoverable
(LRC) codes~\cite{ghsy,pd,tb} by considering them as
erasure-correcting codes. In general II codes are not optimal as LRC
codes with respect to the minimum
distance as the codes in~\cite{tb}. However, II codes require a
much smaller field and they are competitive when metrics
different from the minimum distance are considered, like for example,
the average number of erasures causing an uncorrectable
pattern~\cite{bh}.

Let us point out that LRC codes have important practical
applications, for example, in the Windows Azure storage~\cite{hsx}
and in HDFS-Xorbas~\cite{sa}.

In this paper, we present a novel
definition of EII codes that generalizes previous
definitions. The new definition does not require that the
minimum distances of the nested codes are decreasing. Although the
relaxing of this requirement seems like a small change, the
consequences are profound. In effect, we will show how the new
definition allows for a natural construction of multiple layer EII
codes (a problem already treated in literature for II
codes~\cite{lxx,z3}). This multiple layer construction allows for a
natural hierarchy of localities in EII codes.

The paper is structured as follows:
in Section~\ref{sec2}, we give the definition of EII codes, which is similar
to the traditional definitions of II~\cite{tk,w} and EII~\cite{b,bh2}
codes. As opposed to previous definitions, no
assumption is made with respect to the nested codes utilized in the
construction. In effect, quite often, it is assumed that the nested
codes are MDS and in particular, (shortened) Reed-Solomon codes or at
least that the minimum distance of the nested codes is decreasing. By
not making this assumption, we will show how to construct EII codes
in a hierarchical way. We also present the
fundamental properties of EII codes according to the new definition.
In particular, we give the erasure-correcting capability of the codes,
which we prove constructively by giving an efficient decoding
algorithm and we illustrate it with examples. We also give the
dimension and the minimum distance of the EII codes defined.

In Section~\ref{sec3}, we define recursively $\ell$-layer EII codes
by applying the definition of Section~\ref{sec2}. In
particular, we notice that 2-layer EII codes correspond to
traditional EII codes and that 3-layer~\cite{z3} and multiple
layer~\cite{lxx} II codes are special cases that
were obtained using the so called connection matrices (related to the
parity-check matrices) of the codes. We illustrate
our recursive construction with several concrete examples.

In Section~\ref{secpc} we give parity-check
matrices for the EII codes defined in Section~\ref{sec2} and for the
$\ell$-layer EII codes defined in Section~\ref{sec3}.
The parity-check matrix of an $\ell$-layer EII code is giving
recursively based on parity-check matrices of lower layer EII codes,
and in particular, we give closed formulae for the parity-check matrices of
2 and 3-layer EII codes. We illustrate the construction by revisiting
the examples of Section~\ref{sec3} and obtaining the corresponding
parity-check matrices.

In Section~\ref{secperf} we present a new parameter, the average
number of erasures to failure (ANETF). We argue that the ANETF is
more important than the minimum distance of the code when the
erasures occur one after the other. We run simulations computing the
ANETF of codes
having the same rate as the examples of 2, 3 and 4-layer codes presented
in~\cite{z3} and in~\cite{lxx}. We tabulate the results
and show that in several cases, we obtain codes with better ANETF and
minimum distance than those in~\cite{z3} and in~\cite{lxx}. We also
observe that by using the parity-check matrices at the decoding, we
can often correct erasure patterns that exceed the erasure-correcting
capability of the codes.

We end the paper by drawing some conclusions and ideas for future
research in Section~\ref{conclusions}. In particular, we observe that
the parity-check matrix of a high layer EII code has low density.

\section{Definition and properties of EII codes}
\label{sec2}

We define $t$-level EII codes 
as follows:

\begin{definition}
\label{defl1}
{\rm
Let
$\{\uzero\}\eq\C_t\subset\C_{t-1}\subset\C_{t-2}\subset\cdots
\subset\C_{0}$ be a sequence of $t+1$ nested
$[n,n-u_i,d_i]$ codes over $GF(q)$ for $0\leq i\leq t-1$, $u_0\geq 0$,
$s_0,s_1,\ldots,s_{t}$ non-negative integers such that $1\leq
m=s_0+s_1+\cdots +s_{t-1}+s_t\,<\,q$ and let
$\al$ be an element of order $\cO(\al)\geq m$ in $GF(q)$.

Define $\C$ as the code of length $(m)(n)$
over $GF(q)$, $m<q$, such that, if

\begin{eqnarray}
\label{hsi}
\hs_i &=&
\sum_{j=i}^{t}s_j\quad {\rm for}\quad 0\leq i\leq t,
\end{eqnarray}
for each $\uc\in\C$,
$\uc\eq (\uc_0,\uc_1,\ldots,\uc_{m-1})$, $\uc_j\in\C_0$ for $0\leq
j\leq m-1$ and

\begin{eqnarray}
\label{eqGP1}
\bigoplus_{j=0}^{m-1}\al^{rj}\uc_j&\in& \C_{i}\;\;{\rm for}\;\; 1\leq
i\leq t\;\;{\rm and}\;\; 0\leq r\leq \hs_{i}-1.
\end{eqnarray}

Then we say that $\C$ is an EII code. If $s_t\eq 0$ we say
that $\C$ is an II code. If

\begin{eqnarray}
\label{tlevel}
t'&\eq &|\{i\;:\;0\leq i\leq t-1\quad{\rm and}\quad s_i\neq 0\}|
,\end{eqnarray}
then we say that $\C$ is a $t'$-level EII code.
\qed
}
\end{definition}

Definition~\ref{defl1} is slightly different to the ones
traditionally given in literature.
Most papers on $t$-level II~\cite{w,lxx,z3} or EII codes~\cite{b} assume
that the nested codes $\C_i$ in Definition~\ref{defl1} are
Reed-Solomon (RS) or shortened RS
codes~\cite{ms}. In the original definitions of 2-level
II~\cite{hapkt} and $t$-level II~\cite{tk}
codes, it is not assumed that the nested codes are MDS, although it
is required that their minimum distances are decreasing, i.e.,
$d_i\,<\,d_{i+1}$ for $0\leq i\leq t-1$. There is no such
assumption in Definition~\ref{defl1}. This subtle difference, though,
will be crucial for the construction of $\ell$-layer EII codes to be
presented in Section~\ref{sec3}.

Another difference with traditional definitions is that generally, it
is assumed that $s_i\geq 1$ for $0\leq i\leq t-1$. Taking $s_i\geq 0$
is not fundamentally different, but it is
convenient, since it allows that two different EII
codes share the same nested codes, and this
allows for an easy necessary and sufficient condition, based on the
$s_i$s of each code, to determine if one of the codes is contained in the
other one. This property will be important for the construction of
parity-check matrices of $\ell$-layer EII codes to be given in
Section~\ref{secpc}.

Let us point out that in the literature on $t$-level II codes, a
2-level II code is often
called an II code~\cite{hapkt} while a $t$-level II code with $t>2$
is called a Generalized Integrated Interleaved (GII) code~\cite{lxx,tk,w,z3}.
Since there is no conceptual difference between the cases $t\eq 2$
and $t>2$, we prefer calling these codes simply $t$-level II codes.

The next theorem
describes the erasure patterns that are guaranteed to be corrected by
$t$-level EII codes. The result is a generalization of Theorem~6
in~\cite{bh2}, where the component codes are MDS codes.

\begin{theorem}
\label{theo2}
{\em
Consider an EII code $\C$ on $t+1$ nested codes $\C_i$
according to Definition~\ref{defl1}. Let $\uc\eq (\uc_0,\uc_1,\ldots,\uc_{m-1})\in\C$ be
a codeword with erasures.
Assume that the vectors $\uc_j$ can be divided into $t+1$ disjoint sets
$S_i$, $0\leq i\leq t$, such that $|S_0|\geq s_0$,
$0\leq |S_i|\leq s_i$ for $1\leq i\leq t$ and
the erasures in each $\uc_j\in S_i$, if they occur in a codeword of
code $\C_i$, are correctable in $\C_i$.
Then, all the erasures in $\uc$ can be corrected.
}
\end{theorem}

\noindent\pf
If $\uc_j\in S_0$, since, by Definition~\ref{defl1}, $\uc_j\in\C_0$,
the erasures in $\uc_j$ can be corrected. So,
we may assume that the $\uc_j$s with erasures are not correctable in
$\C_0$, hence, they are not in $S_0$.

Assume that there are $\ell\eq |S_1|+|S_2|+\cdots +|S_t|$ $\uc_j$s that are uncorrectable in $\C_0$
and the erasures in any $\uc_j\in S_i$, 
$1\leq i\leq t$, are
correctable in $\C_i$ when they occur in a codeword of $\C_i$.
We do induction on $\ell$.

If $\ell=0$, there are no erasures and all the sets $S_i$ are empty.
Assume that 
$\ell\geq 1$ 
and the erasures in any $\uc_j\in S_i$ are
correctable in $\C_i$ when they occur in a codeword of $\C_i$.
In particular, $\ell\leq s_1+s_2+\ldots +s_{t}\eq\hs_1$.

Let $i_0,i_1,\ldots,i_{m-1}$
be an ordering of the $\uc_j$s
such that:

\begin{enumerate}

\item If the erasures in $\uc_{i_{\ell-1}}$ are correctable in $\C_w$
when they
occur in a codeword of code $\C_w$, $1\leq w\leq t$, but not
in $\C_{w-1}$ when they occur in a codeword of code $\C_{w-1}$, then the erasures in
$\uc_{i_j}$ for $0\leq j\leq\ell-2$ are not correctable either in
$\C_{w-1}$ when they occur in a codeword of code $\C_{w-1}$.

\item Vectors $\uc_{i_{\ell}},\uc_{i_{\ell+1}},\ldots,\uc_{i_{m-1}}$ have no erasures.

\end{enumerate}

In particular, by 1) and 2) in the ordering of the $\uc_j$s,
$|S_i|\eq 0$ for $1\leq i\leq w-1$ and hence

\begin{eqnarray}
\label{bell}
\ell\eq
|S_w|+|S_{w+1}|+\cdots +|S_t|&\leq &s_w+s_{w+1}+\cdots +s_t\eq\hs_w.
\end{eqnarray}

Rearranging the order of the elements of the sums
in~(\ref{eqGP1}), by~(\ref{bell}), we have

\begin{eqnarray}
\label{eqGP11}
\bigoplus_{j=0}^{m-1}\,\al^{ri_j}\,\uc_{i_j}
&\in& \C_{w}\;\;{\rm for}\;\; 0\leq r\leq \ell -1\leq\hs_w-1.
\end{eqnarray}
Since the $\ell\times m$
matrix corresponding to the
coefficients of the $\uc_{i_j}$s
in~(\ref{eqGP11}) is a Vandermonde type of matrix and
$\cO(\al)\geq m$, this matrix can be triangulated.
Taking the last row of the triangulation, 
we obtain

\begin{eqnarray}
\label{eqGP11tl}
\uc'_{i_{\ell -1}}\;\eq\;\uc_{i_{\ell -1}}
\xor
\left(\bigoplus_{j=\ell}^{m-1}\,\ga_{j}\,\uc_{i_j}\right)
&\in& \C_{w},
\end{eqnarray}
where the coefficients $\ga_j$ are obtained from the triangulation.
Since $\uc_{i_j}$ is erasure free for $\ell\leq j\leq m-1$, then,
by~(\ref{eqGP11tl}), the erasures of $\uc'_{i_{\ell -1}}$ and of
$\uc_{i_{\ell -1}}$ occur in the same locations. Since $\uc'_{i_{\ell
-1}}$ is in $\C_w$ by~(\ref{eqGP11tl}), by condition 1) of the
ordering of the $\uc_j$s, the erasures can be corrected.
Once $\uc'_{i_{\ell -1}}$ is corrected, again by~(\ref{eqGP11tl}),
$\uc_{i_{\ell -1}}$ is obtained as

\begin{eqnarray*}
\uc_{i_{\ell -1}}&=&
\uc'_{i_{\ell -1}}
\xor\left(\bigoplus_{j=\ell}^{m-1}\ga_{j}\,\uc_{i_j}\right).
\end{eqnarray*}

Now we have $\ell-1$ $\uc_j$s with erasures. Redefining $S_0\eq
S_0\cup\{i_{\ell-1}\}$ and $S_w\eq S_w-\{i_{\ell-1}\}$,
the $m$ $\uc_j$s can be divided into $t+1$ disjoint sets
$S_i$, $0\leq i\leq t$, such that $|S_0|\geq s_0$,
$0\leq |S_i|\leq s_i$ for $1\leq i\leq t$ and
the erasures in $\uc_j\in S_i$, if they occur in a codeword of
code $\C_i$, are correctable in $\C_i$.
By induction, the $\ell-1$ $\uc_j$s with erasures can be corrected.
\qed

\begin{corollary}
\label{cormain}
{\em
Consider an EII code $\C$ on $t+1$ nested codes $\C_i$ as in
Definition~\ref{defl1}. Let $\uc\eq
(\uc_0,\uc_1,\ldots,\uc_{m-1})\in\C$ be a codeword with erasures.
Assume that the $m$ $\uc_j$s can be divided into $t+1$ disjoint sets
$S_i$, $0\leq i\leq t$, such that $|S_0|\geq s_0$,
$0\leq |S_i|\leq s_i$ for $1\leq i\leq t$ and each
$\uc_j\in S_i$ has up to $d_i-1$ erasures. 
Then, all the erasures in $\uc$ can be corrected.
In particular, if the codes $\C_i$ are $[n,n-u_i,u_i+1]$ MDS codes,
the code can correct up to $u_i$ erasures in any $\uc_j\in S_i$,
where $0\leq i\leq t$.
}
\end{corollary}
\pf Simply observe that any $d_i-1$ erasures can be corrected in any
codeword of code $\C_i$ for $0\leq i\leq t$ and the result follows
from Theorem~\ref{theo2}. \qed

The MDS case in Corollary~\ref{cormain} corresponds to Theorem~6
in~\cite{bh2}.

Given an EII code $\C$ as in Definition~\ref{defl1}, it is
often convenient to visualize each codeword
$\uc\eq (\uc_0,\uc_1,\ldots,\uc_{m-1})\in\C$ as an $m\times n$ array where
the $\uc_j$s are the rows of the array. We will use indistinctly the
vector and the array description of a codeword in the next examples.

The proof of Theorem~\ref{theo2} provides a
recursive erasure decoding algorithm for $t$-level EII codes, which
we illustrate in Example~\ref{ex11}.

\begin{example}
\label{ex11}
{\em
Assume that
$\{\uzero\}\eq\C_4\subset\C_3\subset\C_2\subset\C_1\subset\C_0$ are
RS codes over $GF(8)$ such that $\C_3$ is a $[7,2]$ code,
$\C_2$ is a $[7,3]$ code, $\C_1$ is a $[7,5]$ code and $\C_0$ is a
$[7,6]$ code. Consider the 4-level EII code $\C$ with $m\eq 7$ and $s_0\eq
2$, $s_1\eq s_2\eq 1$, $s_3\eq 2$ and $s_4\eq 1$ according to
Definition~\ref{defl1}.

Assume that the following $7\times 7$ array is received, where the
entries with $E$ are erased and the blank entries are correct:

$$
\begin{array}{c|c|c|c|c|c|c|c|}
\cline{2-8}
\uc_0&&E&&E&E&E&E\\
\cline{2-8}
\uc_1&E&E&E&E&E&E&E\\
\cline{2-8}
\uc_2&&&E&&&&\\
\cline{2-8}
\uc_3&&E&&E&&E&E\\
\cline{2-8}
\uc_4&E&E&&E&E&&E\\
\cline{2-8}
\uc_5&&&&&&E&\\
\cline{2-8}
\uc_6&&&&E&&E&\\
\cline{2-8}
\end{array}
$$

Dividing the rows in disjoint sets as in Theorem~\ref{theo2}, we
have, $S_0\eq\{2,5\}$, $S_1\eq\{6\}$, $S_2\eq\{3\}$, $S_3\eq\{0,4\}$
and $S_4\eq\{1\}$. Since $|S_0|\eq 2\geq s_0$,
$|S_1|\eq |S_2|\eq 1\leq s_1\eq s_2$, $|S_3|\eq 2\leq s_3$ and
$|S_4|\eq 1\leq s_4$, according to Corollary~\ref{cormain}, the
erasures are correctable. Notice that $\ell\eq
|S_1|+|S_2|+|S_3|+|S_4|\eq 5$.

The first step of the decoding algorithm is correcting the rows with
one erasure, i.e., rows $\uc_2$ and $\uc_5$, so we may assume that
these two rows are erasure-free. Next we reorder the rows according
to the order given in the proof of Theorem~\ref{theo2}, which in this
case would correspond to a non-increasing number of erasures. This
gives, $i_0\eq 1$, $i_1\eq 0$, $i_2\eq 4$, $i_3\eq 3$, $i_4\eq 6$,
$i_5\eq 2$ and $i_6\eq 5$.

Rearranging the order of the elements of the sums
in~(\ref{eqGP1}) as in~(\ref{eqGP11}), since $w$, as defined in
Theorem~\ref{theo2}, is equal to 1 and $\al$ is primitive in $GF(8)$,
thus $\al^7\eq 1$, we obtain

%

\begin{eqnarray*}
\al^4\uc_{1}\xor\uc_0\xor\al^{2}\uc_4\xor\al^{5}\uc_3\xor\al^{3}\uc_6\xor\al
\uc_2\xor\al^{6}\uc_5&\in&\C_4\eq\{\uzero\}\\
\al^3\uc_{1}\xor\uc_0\xor\al^{5}\uc_4\xor\al^{2}\uc_3\xor\al^{4}\uc_6\xor\al^6\uc_2\xor\al
\uc_5&\in&\C_3\\
\al^2\uc_{1}\xor\uc_0\xor\al \uc_4\xor\al^{6}\uc_3\xor\al^{5}\uc_6\xor\al^4\uc_2\xor\al^{3}\uc_5&\in&
\C_3\\
\al\uc_{1}\xor\uc_0\xor\al^4\uc_4\xor\al^3\uc_3\xor\al^6\uc_6\xor\al^2\uc_2\xor\al^5\uc_5&\in& \C_2\\
\uc_{1}\xor\uc_0\xor\uc_4\xor\uc_3\xor\uc_6\xor\uc_2\xor\uc_5&\in& \C_1\\
\end{eqnarray*}

Triangulating above and assuming $\al^3\eq 1+\al$, we obtain

\begin{eqnarray*}
\uc_{1}\xor\al^3\uc_0\xor\al^{5}\uc_4\xor\al\uc_3\xor\al^{6}\uc_6\xor\al^4
\uc_2\xor\al^{2}\uc_5&=&\uzero\\
\uc_0\xor\al^{4}\uc_4\xor\al^{6}\uc_3\xor\al^{6}\uc_6\xor\uc_2\xor\al^4
\uc_5&\in&\C_3\\
\uc_4\xor\al^{6}\uc_3\xor\al^{4}\uc_6\xor\al^6\uc_2\xor\al^{5}\uc_5&\in&
\C_3\\
\uc_3\xor\al^6\uc_6\xor\al^3\uc_2\xor\al^5\uc_5&\in& \C_2\\
\uc_6\xor\al^2\uc_2\xor\al^5\uc_5&\in& \C_1\\
\end{eqnarray*}

Since $\uc_2$ and $\uc_5$ are erasure free, the two erasures of
$\uc_6$ and of $\uc'_6\eq\uc_6\xor\al^2\uc_2\xor\al^5\uc_5$ occur in
the same locations. Since $\uc'_6\in\C_1$ and $\C_1$ is a
[7,5,3] code, these two erasures in $\uc'_6$ can be corrected. Once the
erasures are corrected, we obtain $\uc_6$ as
$\uc_6\eq\uc'_6\xor\al^2\uc_2\xor\al^5\uc_5$.

Similarly, $\uc'_3\eq\uc_3\xor\al^6\uc_6\xor\al^3\uc_2\xor\al^5\uc_5$
and $\uc_3$ have both 4 erasures in the same locations. Since
$\uc'_3\in\C_2$ and $\C_2$ is a $[7,3,5]$ code, the erasures in
$\uc'_3$ are corrected, and then
$\uc_3\eq\uc'_3\xor\al^6\uc_6\xor\al^3\uc_2\xor\al^5\uc_5$.

Next, since
$\uc'_4\eq\uc_4\xor\al^{6}\uc_3\xor\al^{4}\uc_6\xor\al^6\uc_2\xor\al^{5}\uc_5$
and $\uc_4$ have both 5 erasures in the same locations and $\uc'_4\in\C_3$, which is a
$[7,2,6]$ code, the 5 erasures in $\uc'_4$ can be corrected and
$\uc_4\eq\uc'_4\xor\al^{6}\uc_3\xor\al^{4}\uc_6\xor\al^6\uc_2\xor\al^{5}\uc_5$.

Similarly, since $\uc'_0\eq\uc_0\xor\al^{4}\uc_4\xor\al^{6}\uc_3\xor\al^{6}\uc_6\xor\uc_2\xor\al^4
\uc_5$, both $\uc'_0$ and $\uc_0$ have 5 erasures in the same
locations and $\uc'_0\in\C_3$. Correcting the 5 erasures in $\uc'_0$, then $\uc_0\eq
\uc'_0\xor\al^{4}\uc_4\xor\al^{6}\uc_3\xor\al^{6}\uc_6\xor\uc_2\xor\al^4 \uc_5$.

Finally, $\uc_1$ is obtained as $\uc_{1}\eq\al^3\uc_0\xor\al^{5}\uc_4\xor\al\uc_3\xor\al^{6}\uc_6\xor\al^4
\uc_2\xor\al^{2}\uc_5$, completing the decoding.
}
\end{example}

In particular,
the encoding is a special case of the decoding.
In effect, without
loss of generality, we may assume that each of the codes $\C_i$,
$0\leq i\leq t$, in Definition~\ref{defl1}, admits a systematic
encoder such that the first $n-u_i$ symbols in a codeword contain
data while the last $u_i$ symbols contain parity. If we view the $u_i$
parity symbols as erasures, then such erasures are correctable by
$\C_i$. Say, we take an $m\times n$ array such that in the first
$s_0$ rows the first $n-u_0$ symbols in each row contain data, in the next
$s_1$ rows the first $n-u_1$ symbols in each row contain data, and so
on, until the last $s_t$ rows, in which all the symbols contain
parity. According to Theorem~\ref{theo2}, the erasures can be solved,
so the recursive algorithm in the theorem can be used as an encoder.
The fact that at the encoding the locations of the erasures
are known allows for a simplification of the decoding algorithm. For
example, the triangulated matrix 
in the proof of Theorem~\ref{theo2} may be precomputed. The next
example illustrates the encoding algorithm.

\begin{example}
\label{ex12}
{\em
Let us retake the 4-level EII code of Example~\ref{ex11}, and
according to the description above, for the encoding we can place the parity and data
as follows:

$$
\begin{array}{c|c|c|c|c|c|c|c|}
\cline{2-8}
\uc_0&D&D&D&D&D&D&P\\
\cline{2-8}
\uc_1&D&D&D&D&D&D&P\\
\cline{2-8}
\uc_2&D&D&D&D&D&P&P\\
\cline{2-8}
\uc_3&D&D&D&P&P&P&P\\
\cline{2-8}
\uc_4&D&D&P&P&P&P&P\\
\cline{2-8}
\uc_5&D&D&P&P&P&P&P\\
\cline{2-8}
\uc_6&P&P&P&P&P&P&P\\
\cline{2-8}
\end{array}
$$
where $D$ denotes data and $P$ denotes parity. As stated, we may
consider the parities as erasures and apply the decoding algorithm to
them.

The disjoint sets as in Theorem~\ref{theo2} are then
$S_0\eq\{0,1\}$, $S_1\eq\{2\}$, $S_2\eq\{3\}$, $S_3\eq\{4,5\}$
and $S_4\eq\{6\}$.

As in Example~\ref{ex11}, we first correct the rows with
one erasure, in this case, rows $\uc_0$ and $\uc_1$, so after doing
so, these two rows are erasure-free. The order of the rows given in
the proof of Theorem~\ref{theo2} is $i_j\eq 6-j$ for $0\leq
j\leq 6$. Then we have,

\begin{eqnarray*}
\al^3\uc_6\xor\al^6\uc_5\xor\al^2\uc_4\xor\al^5\uc_3\xor\al\uc_2\xor\al^4\uc_1\xor\uc_{0}&=&\uzero\\
\al^4\uc_6\xor\al\uc_5\xor\al^5\uc_4\xor\al^2\uc_3\xor\al^6\uc_2\xor\al^3\uc_1\xor\uc_{0}&\in&
\C_3\\
\al^5\uc_6\xor\al^3\uc_5\xor\al\uc_4\xor\al^6\uc_3\xor\al^4\uc_2\xor\al^2\uc_1\xor\uc_{0}&\in&
\C_3\\
\al^6\uc_6\xor\al^5\uc_5\xor\al^4\uc_4\xor\al^3\uc_3\xor\al^2\uc_2\xor\al\uc_1\xor\uc_{0}&\in& \C_2\\
\uc_{6}\xor\uc_5\xor\uc_4\xor\uc_3\xor\uc_2\xor\uc_1\xor\uc_0&\in& \C_1\\
\end{eqnarray*}

Triangulating above and assuming $\al^3\eq 1+\al$, we obtain

\begin{eqnarray*}
\uc_{6}\xor\al^3\uc_5\xor\al^6\uc_4\xor\al^2\uc_3\xor\al^5\uc_2\xor\al\uc_1\xor\al^4\uc_0&=&\uzero\\
\uc_5\xor\al^6\uc_4\xor\al^4\uc_3\xor\al^4\uc_2\xor\al^6\uc_1\xor\uc_0&\in&\C_3\\
\uc_4\xor\al\uc_3\xor\al^3\uc_2\xor\al^2\uc_1\xor\al^2\uc_0&\in&\C_3\\
\uc_3\xor\al^5\uc_2\xor\al^6\uc_1\xor\al^3\uc_0&\in& \C_2\\
\uc_2\xor\al^4\uc_1\xor\al^3\uc_0&\in& \C_1\\
\end{eqnarray*}

Since at the encoding we know the location of the erasures, this
triangulated matrix can be precomputed. We then obtain successively
$\uc_2$, $\uc_3$, $\uc_4$, $\uc_5$ and $\uc_6$ as in
Example~\ref{ex11}, completing the encoding.
}
\end{example}

Since the $mn-\sum_{i=0}^ts_iu_i$ data symbols at the encoding are completely
arbitrary, we have the following theorem:

\begin{theorem}
\label{cor00}
{\em
Consider an EII code $\C$ as given by
Definition~\ref{defl1}. Then, $\C$ is an $[(m)(n),k]$ code, where

\begin{eqnarray}
\label{eqK}
k&\,=\,&
(m)(n)-\left(\sum_{i=0}^{t}s_iu_i\right).
\end{eqnarray}
\qed
}
\end{theorem}

Theorem~\ref{cor00} coincides with Theorem~12 in~\cite{bh2}, the only
difference being that the nested codes in~\cite{bh2} are RS or
shortened RS type of codes, while no such limitation is required in
Theorem~\ref{cor00}.

\begin{example}
\label{ex110}
{\em
Let $\C$ be the 4-level EII code of Example~\ref{ex11}. Then, $\C$ is a
$[49,k]$ code where, according to~(\ref{eqK}),

\begin{eqnarray*}
k&\eq &49-(2)(1)-(1)(2)-(1)(3)-(2)(5)-(1)(7)\quad\eq \quad 25.
\end{eqnarray*}
}
\end{example}

The next theorem gives the minimum distance of an
EII code.

\begin{theorem}
\label{lemma1}
{\em
Consider an 
EII code $\C$ as given by Definition~\ref{defl1}.
Then,

\begin{eqnarray}
\label{ubdist}
d&\eq &\min\left\{d_j\left(\hs_{j+1}+1\right)\;\;{\rm for}\;\;0\leq
j\leq t-1\right\}
\end{eqnarray}
}
\end{theorem}

\noindent\pf
Take $j$ such that $0\leq j\leq t-1$,
We prove that there is a codeword of weight $d_j\left(\hs_{j+1}+1\right)$.

Since $\C_j$ is an $[n,n-u_j,d_j]$ code, 
there is a codeword $\uw\in\C_j$ of weight $d_j$.


Consider the polynomial $\uv(x)\eq (x\xor 1)(x\xor\al)\ldots
(x\xor\al^{\hs_{j+1}-1})\eq v_0+v_1x+\cdots
+v_{\hs_{j+1}}x^{\hs_{j+1}}$. In particular, $v_s\neq 0$ for
$0\leq s\leq \hs_{j+1}$ and

\begin{eqnarray}
\label{ref}
\uv(\al^r)\quad \eq\quad\bigoplus_{s=0}^{\hs_{j+1}}\al^{rs}v_s&=&0\;\;{\rm for}\;\;0\leq
r\leq \hs_{j+1}-1.
\end{eqnarray}

Take a vector $\uc\eq (\uc_0,\uc_1,\ldots,\uc_{m-1})$
such that 
$\uc_s\eq v_s\,\uw$ for $0\leq s\leq \hs_{j+1}$ and $\uc_s\eq\uzero$
for $\hs_{j+1}+1\leq s\leq m-1$.
In particular, $\uc$ has weight $d_j\left(\hs_{j+1}+1\right)$ and
we will show that $\uc\in\C$ . Since 
$\uc_s\in\C_j$ by design and $\C_j\subseteq\C_0$, in particular,
$\uc_s\in\C_0$ for $0\leq s\leq m-1$.
According to~(\ref{eqGP1}), we also have to show that

\begin{eqnarray}
\label{eqGP11b}
\bigoplus_{s=0}^{m-1}\al^{rs}\uc_s\quad
=\quad\bigoplus_{s=0}^{\hs_{j+1}}\al^{rs}\left(v_s\,\uw\right)&\in&
\C_{i}\;\;{\rm for}\;\; 1\leq i\leq t\;\;
{\rm and}\;\;0\leq r\leq \hs_{i}-1.
\end{eqnarray}

Take $i$ such that $1\leq i\leq t$ and $r$ such that $0\leq r\leq \hs_i-1$.
Assume first that $j<i$, then
$j+1\leq i$ and, by~(\ref{hsi}), $\hs_{j+1}\geq \hs_i$, so, in
particular, $0\leq r\leq\hs_{j+1}-1$. Then, by~(\ref{ref}),

\begin{eqnarray*}
\bigoplus_{s=0}^{\hs_{j+1}}\al^{rs}\left(v_s\,\uw\right)\quad\eq\quad
\left(\bigoplus_{s=0}^{\hs_{j+1}}\al^{rs}v_s\right)\,\uw\quad\eq\quad\uzero
\end{eqnarray*}
and in particular, (\ref{eqGP11b}) holds.

Assume next that $j\geq i$, then, $\C_j\subseteq\C_i$. Hence, since
$\uw\in\C_j$, also $\uw\in\C_i$ and~(\ref{eqGP11b}) holds, so
$d\leq d_j\left(\hs_{j+1}+1\right)$. In particular,
$d\leq \min\left\{d_j\left(\hs_{j+1}+1\right)\;\; {\rm for}\;\; 0\leq
j\leq t-1\right\}$.

We prove the other inequality next.
Assume that $d<\min\left\{d_j\left(\hs_{j+1}+1\right)\;\;{\rm for}\;\;0\leq
j\leq t-1\right\}$ and take a codeword  of weight $d$
$\uc\eq (\uc_0,\uc_1,\ldots,\uc_{m-1})\in\C$.
Assume that the $d$ non-zero entries of $\uc$ are erased.
Let $\uc_{i_0},\uc_{i_1},\ldots,\uc_{i_{\ell-1}},\uc_{i_{\ell}},\ldots,\uc_{i_{m-1}}$
be the vectors of $\uc$ ordered in non-increasing weight order and assume
that vectors $i_0$ to $i_{\ell-1}$ have non-zero weight, i.e., if
$\uc_{i_s}$ has weight $w_s$, then $\sum_{s=0}^{\ell-1}w_s\eq d$ and
$w_0\geq w_1\geq\cdots\geq w_{\ell-1}$.
If $w_{\ell-1}<d_0$, vector $\uc_{i_{\ell-1}}$ would be corrected in $\C_0$ as the
zero vector, contradicting that $w_{\ell-1}\neq 0$, so $w_{\ell-1}\geq
d_0$. Also, $\ell>s_t$, otherwise $\uc$ would be corrected as the zero
array by Theorem~\ref{theo2}. Hence, we can define
$i$, $1\leq i\leq t-1$, such that $\hs_{i+1}<\ell\leq \hs_i$. Assume
that $w_{\ell-1}\geq d_i$. Then,

\begin{eqnarray*}
d\quad\eq\quad \sum_{s=0}^{\ell-1}w_s\quad\geq\quad w_{\ell-1}\ell\quad\geq\quad d_i(\hs_{i+1}+1),
\end{eqnarray*}
contradicting the assumption that $d<\min\left\{d_j\left(\hs_{j+1}+1\right)\;\;{\rm for}\;\;0\leq
j\leq t-1\right\}$. Then, $w_{\ell-1}<d_i$.
Since $\uc\in\C$, $\uc_{i_s}\eq\uzero$ for $\ell\leq s\leq m-1$ and
$\ell\leq \hs_i$, rearranging the order of the elements of the sums
in~(\ref{eqGP1}), we obtain

\begin{eqnarray}
\label{eqGP111}
\bigoplus_{s=0}^{m-1}\,\al^{ri_s}\,\uc_{i_s}\quad\eq\quad\bigoplus_{s=0}^{\ell-1}\,\al^{ri_s}\,\uc_{i_s}
&\in& \C_{i}\;\;{\rm for}\;\; 0\leq r\leq \ell -1.
\end{eqnarray}

Since the $\ell\times \ell$ matrix corresponding to the
coefficients of the $\uc_{i_s}$s in~(\ref{eqGP111})
is a Vandermonde type of matrix and
$\cO(\al)\geq m$, this matrix can be triangulated and
$\uc_{i_{\ell -1}}\in\C_i$. Since $\C_i$ has minimum distance
$d_i$ and $\uc_{i_{\ell -1}}$ has weight $w_{\ell-1}<d_i$, then
$\uc_{i_{\ell -1}}\eq\uzero$, a contradiction.
\qed

The following corollary corresponds to Theorem~15 in~\cite{bh2}.

\begin{corollary}
\label{immediate}
{\em
Consider a $t$-level EII code $\C$ as given
by~Definition~\ref{defl1} such that, for $0\leq j\leq t-1$, code
$\C_j$ is an $[n,n-u_j,u_j+1]$ MDS code.
Then, the minimum distance of $\C$ is

\begin{eqnarray}
\label{distMDS}
d&=&\min\left\{\left(u_j+1\right)\left(\hs_{j+1}+1\right)\;\;{\rm for}\;\;0\leq
j\leq t-1\right\}.
\end{eqnarray}
}
\end{corollary}

\noindent\pf
Simply notice that $d_j\eq u_j+1$ for $0\leq
j\leq t-1$ and~(\ref{ubdist}) gives~(\ref{distMDS}).
\qed

\begin{example}
\label{ex111}
{\em
Let $\C$ be again the 4-level EII code of Example~\ref{ex11}. Then, according to~(\ref{distMDS}),

\begin{eqnarray*}
d&=&\min\left\{(2)(6)\,,\,(3)(5)\,,\,(5)(4)\,,\,(6)(2)\right\}\quad\eq\quad 12.
\end{eqnarray*}
}
\end{example}

We end this section with a lemma providing necessary and sufficient
conditions to determine whether, given two EII codes
sharing the same nested codes, one of them is
contained in the other. 

\begin{lemma}
\label{lemma2}
{\em
Let $\C$ and $\C'$ be two
EII codes with the same nested nested codes $\{\uzero\}\eq\C_{t}
\subset\C_{t-1}\subset\C_{t-2}\subset\cdots\subset
\C_{0}$ and non-negative coefficients $s_i$ and $s'_i$ respectively
for $0\leq i\leq t$
according to Definition~\ref{defl1}. Then,
$\C'\subseteq \C$ if and only if
$\hs_{i}\leq\hs'_{i}$ for $0\leq i\leq t$.
}
\end{lemma}

\noindent\pf
Let $\uc\eq
(\uc_0,\uc_1,\ldots,\uc_{m-1})\in\C'$.
We have to prove that $\uc\in\C$ if and only if
$\hs_{i}\leq\hs'_{i}$ for $0\leq i\leq t$.

Since $\uc\in\C'$, by
Definition~\ref{defl1}, $\uc_j\in\C_{0}$ for
$0\leq j\leq m-1$ and, by~(\ref{eqGP1}),

\begin{eqnarray}
\label{eqGP2bis}
\bigoplus_{j=0}^{m-1}\al^{rj}\uc^{(\ell-1)}_j&\in& \C_{i}\;\;{\rm for}\;\; 1\leq
i\leq t\;\;{\rm and}\;\; 0\leq r\leq \hs'_{i}-1.
\end{eqnarray}

Then, $\uc\in\C$ if and only if (\ref{eqGP1})
holds, which, by~(\ref{eqGP2bis}), will be the case if and only if $\hs_{i}\leq\hs'_{i}$ for $0\leq i\leq t$.
%
\qed

Lemma~\ref{lemma2} will be useful when constructing the parity-check
matrices of $\ell$-layer EII codes to be presented in Section~\ref{secpc}.

\section{Recursive construction of $\ell$-layer EII codes}
\label{sec3}
The concept of 3-layer II codes is presented in~\cite{z} and its
generalization to multi-layer II codes in~\cite{lxx}. Next we are
going to show that these concepts arise naturally by applying
recursively Definition~\ref{defl1} of EII
codes, as shown in the next definition. We also automatically obtain
the properties of $\ell$-layer EII codes discussed in
Section~\ref{sec2}, like their dimension
and minimum distance.

\begin{definition}
\label{defmlayer}
{\rm We say that $\C^{(1)}$ is a 1-layer EII code if it is an $[n,n-u,u+1]$ code over
$GF(q)$.
Assuming that $\ell$-layer EII codes
of length $(m')(n)$ over $GF(q)$ have been defined for $\ell\geq 1$, where
$m_0\eq 1$ and $m'\eq (m_{\ell-1})(m_{\ell-2})\ldots (m_1)(m_0)$, let
$\{\uzero\}\eq\C^{(\ell)}_{t}
\subset\C^{(\ell)}_{t-1}\subset\C^{(\ell)}_{t-2}\subset\cdots\subset \C^{(\ell)}_{0}$
be a sequence of $t+1$ nested $\ell$-layer EII codes, $s_0+s_1+\cdots +s_t\eq
m_{\ell}$, $s_i\geq 0$ for $0\leq i\leq t$, and $\al\in GF(q)$ such that $\cO(\al)\geq m_{\ell}$. Then,
we say that $\C^{(\ell+1)}$ is an $(\ell+1)$-layer
EII code of length $(m_{\ell})(m')(n)$ if $\C^{(\ell+1)}$ is an EII
code over the nested $\ell$-layer codes $\C^{(\ell)}_i$ according to Definition~\ref{defl1}.

If $s_t\eq 0$, we say that $\C^{(\ell+1)}$ is an $(\ell+1)$-layer II code.
\qed
}
\end{definition}

Comparing Definitions~\ref{defl1} and~\ref{defmlayer}, we see that an
$(\ell+1)$-layer code is an EII code such that the nested
codes are $\ell$-layer EII codes.
Hence,
a 1-layer EII code is simply
an MDS code while a 2-layer EII code corresponds to the
EII code of Definition~\ref{defl1} such that the nested codes are MDS
codes (this assumption is made in most papers on II
codes~\cite{bh3,bh,lxx,w,z3}).  

Let us point out also that although not required in
Definition~\ref{defmlayer}, it is convenient to use a unique element
$\al$ in all the layers by requiring
$\cO(\al)\geq\max\{m_1,m_2,\ldots,m_{\ell}\}$.

Next we define recursively the
erasure-correcting capability of $\ell$-layer EII codes by using a
vector $\uu$.

\begin{definition}
\label{ecap}
{\rm If $\C^{(1)}$ is
a 1-layer $[n,n-u,u+1]$ EII code, we say that the erasure-correcting capability
of $\C^{(1)}$ is the vector of length 1 $\uu^{(1)}\eq (u)$.
Let $\ell\geq 1$ and consider an
$(\ell+1)$-layer EII code $\C^{(\ell+1)}$ as given by
Definition~\ref{defmlayer}. Let the erasure-correcting capability of
the $\ell$-layer nested EII code				
$\C^{(\ell)}_i$, $0\leq i\leq t-1$, be given by a vector $\uu^{(\ell)}_i$ of length
$(m_{\ell-1})(m_{\ell-2})\ldots (m_1)(m_0)$, where $m_0\eq 1$.
Then, we denote the erasure-correcting
capability of $\C^{(\ell+1)}$ by the vector of length
$(m_{\ell})(m_{\ell-1})\ldots (m_1)(m_0)$

\begin{eqnarray}
\label{uu}
\uu^{(\ell+1)}\eq
\left(\overbrace{\uu_0^{(\ell)},\uu_0^{(\ell)},\ldots,\uu_0^{(\ell)}}^{s_0},
\overbrace{\uu_1^{(\ell)},\uu_1^{(\ell)},\ldots,\uu_1^{(\ell)}}^{s_1},\ldots,
\overbrace{\uu_{t-1}^{(\ell)},\uu_{t-1}^{(\ell)},\ldots,\uu_{t-1}^{(\ell)}}^{s_{t-1}}\right).
\end{eqnarray}
\qed
}
\end{definition}





We illustrate Definitions~\ref{defmlayer} and~\ref{ecap} in the next examples.

\begin{example}
\label{ex00}
{\em
Let $n\eq 7$, $m_1\eq 6$, $\C^{(1)}_i$ a $[7,7-i-1,i+2]$ MDS code over
$GF(8)$ for $0\leq i\leq 6$ and $\al$ a primitive element in $GF(8)$.
Let $\C^{(2)}_0$ and $\C^{(2)}_1$ be the two 2-layer 2-level
II codes with nested codes
$\{\uzero_7\}\eq\C^{(1)}_2\subset\C^{(1)}_1\subset\C^{(1)}_0$. Denoting by
$s_{i,j}$ the $s_j$s of code $\C{(2)}_i$, $0\leq i\leq 1$, according to
Definition~\ref{defmlayer}, assume that
$s_{0,0}\eq 5$, $s_{0,1}\eq 1$, $s_{0,2}\eq 0$, $s_{1,0}\eq 4$,
$s_{1,1}\eq 2$ and $s_{0,2}\eq 0$.
By Lemma~\ref{lemma2}, $\C^{(2)}_1\subset\C^{(2)}_0$.
By Theorem~\ref{theo2}, both codes can correct any of the 6 rows with
one erasure, and in
addition, $\C^{(2)}_0$ can correct up to one row with two erasures,
while $\C^{(2)}_1$ can correct any pair of rows with two erasures each.
By~(\ref{uu}) in Definition~\ref{ecap}, the erasure-correcting capability of $\C^{(2)}_0$ is
$\uu_0^{(2)}\eq (1,1,1,1,1,2)$ and the erasure-correcting capability
of $\C^{(2)}_1$ is $\uu_1^{(2)}\eq (1,1,1,1,2,2)$.
By Theorem~\ref{cor00} and Corollary~\ref{immediate},
$\C^{(2)}_0$ is a $[42,35,3]$ code and $\C^{(2)}_1$ is a $[42,34,3]$
code.

Since $\{\uzero_{42}\}\eq\C^{(2)}_2\subset \C^{(2)}_1\subset
\C^{(2)}_0$ we can construct a 3-layer
2-level II code $\C^{(3)}$ using Definition~\ref{defmlayer}.
We note here that both nested codes $\C^{(2)}_0$ and $\C^{(2)}_1$
have the same minimum distance $d_0\eq d_1\eq 3$, while the traditional
definition of II codes~\cite{hapkt,tk} requires $d_0<d_1$. Dropping
this requirement allows us to construct higher layer EII codes.

In effect,
assume that $m_2\eq 2$, $s_0\eq s_1\eq 1$ and $s_2\eq 0$. Then, if
$\uc^{(3)}\in\C^{(3)}$, $\uc^{(3)}\eq (\uc^{(2)}_0\,,\,\uc^{(2)}_1)$,
where $\uc^{(2)}_0,\uc^{(2)}_1\in\C^{(2)}_0$ and, according
to~(\ref{eqGP1}), $\uc^{(2)}_0\xor\uc^{(2)}_1\in\C^{(2)}_1$.

We can visualize both $\uc^{(2)}_0$ and $\uc^{(2)}_1$ as $6\times 7$ arrays.
Then, according to
Theorem~\ref{theo2}, code $\C^{(3)}$ can correct an array with
erasures correctable in $\C^{(2)}_0$ together with an array with
erasures correctable in $\C^{(2)}_1$.
For example, consider the two arrays in $\C^{(3)}$

$$
\begin{array}{cc}
\begin{array}{|c|c|c|c|c|c|c|}
\hline
\phantom{E}&E&\phantom{E}&\phantom{E}&\phantom{E}&\phantom{E}&\phantom{E}\\
\hline
E&&&&&&\\
\hline
&&E&E&&&\\
\hline
&&&&&&E\\
\hline
&&&E&&E&\\
\hline
&&&&E&&\\
\hline
\end{array}
&
\begin{array}{|c|c|c|c|c|c|c|}
\hline
\phantom{E}&\phantom{E}&E&\phantom{E}&\phantom{E}&\phantom{E}&\phantom{E}\\
\hline
&&E&&&&\\
\hline
&E&&&&E&\\
\hline
&&&&&&E\\
\hline
&&&E&&&\\
\hline
&&&&E&&\\
\hline
\end{array}
\end{array}
$$
where $E$ denotes an erasure. Each row with only one erasure is in
$\C^{(1)}_0$, so it can be corrected. After correcting the rows with
one erasure, since each of the two arrays is
in $\C^{(2)}_0$, the array in the right, which has one row with
two erasures while the remaining ones are erasure-free, is corrected
without intervention of the first array.
Once this array is corrected, the array
in the left, which has two rows with two erasures and the remaining
ones are erasure-free, is corrected
following the method of Theorem~\ref{theo2}.

By~(\ref{uu}) in Definition~\ref{ecap}, the erasure-correcting capability of code $\C^{(3)}$
is $\uu^{(3)}\eq ((1,1,1,1,1,2),(1,1,1,1,2,2))$.

By Theorems~\ref{cor00} and~\ref{lemma1}, code $\C^{(3)}$ is an $[84,69,3]$ code.
}
\end{example}

\begin{example}
\label{ex01}
{\em
This example is similar to the example in Section~IV of~\cite{lxx}.
Let $n\eq 7$, $\C^{(1)}_i$ 1-layer EII codes over $GF(8)$ as in
Example~\ref{ex00} and $\al$ a primitive
element in $GF(8)$. Let $m_1\eq 3$ and
$\C^{(2)}_0$ and $\C^{(2)}_1$ two 2-layer EII codes with nested codes
$\{\uzero_7\}\eq\C^{(1)}_3\subset\C^{(1)}_2\subset\C^{(1)}_1\subset\C^{(1)}_0$. As in
Example~\ref{ex00}, we denote by $s_{i,j}$ the
$s_j$s of $\C^{(2)}_i$, $0\leq i\leq 1$, according to
Definition~\ref{defmlayer}.

Assume that
$s_{0,0}\eq 2$, $s_{0,1}\eq 1$, $s_{0,2}\eq 0$, $s_{0,3}\eq 0$,
$s_{1,0}\eq 1$, $s_{1,1}\eq 1$, $s_{1,2}\eq 1$ and $s_{1,3}\eq 0$.
Notice that $\C^{(2)}_0$ is a 2-layer 2-level II code while
$\C^{(2)}_1$ is a 2-layer 3-level II code.
By Theorem~\ref{cor00} and Corollary~\ref{immediate},
$\C^{(2)}_0$ is a $[21,17,3]$ code and $\C^{(2)}_1$ is a $[21,15,4]$
code. 
Considering each codeword
as a $3\times 7$ array, by Theorem~\ref{theo2}, both codes can
correct any of the three rows with
one erasure, and in
addition, $\C^{(2)}_0$ can correct up to one row with two erasures,
and $\C^{(2)}_1$ can correct one row with two erasures and one row
with three erasures.
By~(\ref{uu}) in Definition~\ref{ecap}, the erasure-correcting capability of $\C^{(2)}_0$ is
$\uu_0^{(2)}\eq (1,1,2)$ and the erasure-correcting capability
of $\C^{(2)}_1$ is $\uu_1^{(2)}\eq (1,2,3)$.

By Lemma~\ref{lemma2}, $\C^{(2)}_1\subset \C^{(2)}_0$ and hence we can construct a 3-layer
II code $\C^{(3)}$ using Definition~\ref{defl1} with nested codes
$\C^{(2)}_0$ and $\C^{(2)}_1$. In effect,
assume that $m_2\eq 4$, $s_0\eq 1$,  $s_1\eq 3$ and  $s_2\eq 0$,
hence, $\C^{(3)}$ is a 3-layer 2-level II code. We may
visualize the codewords in $\C^{(3)}$ as four $3\times 7$ arrays. Then, according to
Theorem~\ref{theo2}, any erasures correctable in $\C^{(2)}_0$ can
be corrected in any of the arrays, while up to three arrays with
erasures correctable in $\C^{(2)}_1$  are also correctable in
$\C^{(3)}$ provided that the fourth array is erasure-free. For
example, consider the four arrays in $\C^{(3)}$

$$
\begin{array}{cccc}
\begin{array}{|c|c|c|c|c|c|c|}
\hline
\phantom{E}&E&\phantom{E}&E&\phantom{E}&E&\phantom{E}\\
\hline
E&&&&&&\\
\hline
&&E&E&&&\\
\hline
\end{array}
&
\begin{array}{|c|c|c|c|c|c|c|}
\hline
\phantom{E}&\phantom{E}&E&\phantom{E}&\phantom{E}&\phantom{E}&\phantom{E}\\
\hline
&E&&&&E&\\
\hline
&&&&&&E\\
\hline
\end{array}
&
\begin{array}{|c|c|c|c|c|c|c|}
\hline
\phantom{E}&\phantom{E}&\phantom{E}&\phantom{E}&\phantom{E}&\phantom{E}&E\\
\hline
&&&E&&E&\\
\hline
E&&&&E&&E\\
\hline
\end{array}
&
\begin{array}{|c|c|c|c|c|c|c|}
\hline
\phantom{E}&E&E&\phantom{E}&\phantom{E}&\phantom{E}&\phantom{E}\\
\hline
&&&E&&&\\
\hline
&E&&&E&&E\\
\hline
\end{array}
\end{array}
$$
Each row with only one erasure is in
$\C^{(1)}_0$, so it can be corrected. After correcting the rows with
one erasure, since the second array is
in $\C^{(2)}_0$ and it has one row with
two erasures while the remaining two are erasure-free, it is corrected.
Once this array is corrected, the three remaining arrays,
which have a row with two erasures, a row with three erasures and the remaining
one is erasure-free, are corrected
following the decoding algorithm of Theorem~\ref{theo2}.

By~(\ref{uu}) in Definition~\ref{ecap}, the erasure-correcting capability of code $\C^{(3)}$
is $\uu^{(3)}\eq ((1,1,2),(1,2,3),(1,2,3),(1,2,3))$.

By Theorems~\ref{cor00} and~\ref{lemma1} code $\C^{(3)}$ is an $[84,62,4]$ code.

As a comparison (and as was done in~\cite{lxx}), consider a 2-layer
2-level code $\C^{(2)}$ with $m\eq 12$, also with nested 1-layer EII codes
$\C_2^{(1)}\subset \C_1^{(1)}\subset \C_0^{(1)}$ as above, and
$s_0\eq 5$, $s_1\eq 4$ and $s_2\eq 3$. However, since $m\eq
12$ and, according to Definition~\ref{defl1}, $m<q$, we cannot use
the field $GF(8)$ as in the case of code $\C^{(3)}$ above. The next
field of characteristic 2 is $GF(16)$, so we assume that the codes
$\C^{(1)}_i$ above are over $GF(16)$.
By Theorem~\ref{cor00} and Corollary~\ref{immediate}, $\C^{(2)}$ is
also an $[84,62,4]$ code, so $\C^{(2)}$ and $\C^{(3)}$ have the same rate and
minimum distance. However,
since by~(\ref{uu}) in Definition~\ref{ecap}, the erasure-correcting
capability of $\C^{(2)}$ is (1,1,1,1,1,2,2,2,2,3,3,3), there
are erasures that can be corrected in $\C^{(2)}$ but not in
$\C^{(3)}$. For example, visualizing the codewords in $\C^{(2)}$ as
$12\times 7$ arrays, $\C^{(2)}$ can correct any
three rows with three erasures each while the remaining nine rows are
erasure-free by Theorem~\ref{theo2}. Certainly this is not true for code
$\C^{(3)}$. But code $\C^{(3)}$ has better locality than code
$\C^{(2)}$: if a row has two erasures and the remaining 11 are
erasure-free, then all such 12 rows are needed to reconstruct the
erasures in $\C^{(2)}$, while $\C^{(3)}$ can do it with only
three rows. In addition, since code $\C^{(3)}$ is over $GF(8)$ while code
$\C^{(2)}$ is over $GF(16)$, the implementation of $\C^{(3)}$ has
less complexity. These are tradeoffs that must be taken into account
when choosing a code.
}
\end{example}

\begin{example}
\label{ex04}
{\em
This example is similar to Example~\ref{ex01}, but we incorporate an
EII code (as opposed to an II code) as one of the nested 2-layer EII
codes (notice that~\cite{lxx,z3} use only II codes as nested codes).

Let $m_1\eq 3$ and
$\C^{(2)}_0$ and $\C'^{(2)}_1$ two 2-layer EII codes with nested
1-layer II codes
$\{\uzero_7\}\eq\C^{(1)}_2\subset\C^{(1)}_1\subset\C^{(1)}_0$, where
$\C^{(1)}_0$ and $\C^{(1)}_1$
are as in Example~\ref{ex01}. We denote by $s_{0,j}$ the
$s_j$s of $\C^{(2)}_0$ and by $s_{1,j}$ the
$s_j$s of $\C'^{(2)}_1$ according to Definition~\ref{defmlayer}.

Assume that
$s_{0,0}\eq 2$, $s_{0,1}\eq 1$, $s_{0,2}\eq 0$,
$s_{1,0}\eq 1$, $s_{1,1}\eq 1$ and $s_{1,2}\eq 1$.
Notice that, since $s_{0,2}\eq 0$, $\C^{(2)}_0$ is a 2-layer 2-level
II code while, since $s_{1,2}\neq 0$,
$\C'^{(2)}_1$ is a 2-layer 2-level EII code.
By Theorem~\ref{cor00} and Corollary~\ref{immediate},
$\C^{(2)}_0$ is a $[21,17,3]$ code  and $\C'^{(2)}_1$ is a $[21,11,6]$
code. By~(\ref{uu}) in Definition~\ref{ecap}, the erasure-correcting
capability of $\C^{(2)}_0$ is $(1,1,2)$ while the one of $\C'^{(2)}_1$
is $(1,2,7)$.
By Lemma~\ref{lemma2}, $\C'^{(2)}_1\subset\C^{(2)}_0$.

Next we construct a 3-layer
2-level II code $\C'^{(3)}$ using Definition~\ref{defl1} on the
nested codes $\{\uzero_{21}\}\eq\C^{(2)}_2\subset\C'^{(2)}_1\subset \C^{(2)}_0$  with $m_2\eq 4$,
$s_0\eq 3$, $s_1\eq 1$ and $s_2\eq 0$. As in Example~\ref{ex01}, we visualize a
codeword in $\C'^{(3)}$ as four $3\times 7$ arrays. According to
Theorem~\ref{theo2}, any erasures correctable in $\C^{(2)}_0$ can
be corrected in any of the arrays, while up to one of the arrays with
erasures correctable in $\C'^{(2)}_1$  is also correctable in
$\C'^{(3)}$ provided that the remaining three arrays are
erasure-free. According to~(\ref{uu}), the erasure-correcting
capability of $\C'^{(3)}$ is
$\uu'^{(3)}\eq\left((1,1,2),(1,1,2),(1,1,2),(1,2,7)\right)$. For
example, consider the four arrays in $\C'^{(3)}$

$$
\begin{array}{cccc}
\begin{array}{|c|c|c|c|c|c|c|}
\hline
\phantom{E}&\phantom{E}&\phantom{E}&\phantom{E}&\phantom{E}&E&\phantom{E}\\
\hline
E&&&&&&\\
\hline
&&E&E&&&\\
\hline
\end{array}
&
\begin{array}{|c|c|c|c|c|c|c|}
\hline
\phantom{E}&\phantom{E}&E&\phantom{E}&\phantom{E}&\phantom{E}&\phantom{E}\\
\hline
&E&&&&E&\\
\hline
&&&&&&E\\
\hline
\end{array}
&
\begin{array}{|c|c|c|c|c|c|c|}
\hline
\phantom{E}&\phantom{E}&\phantom{E}&\phantom{E}&\phantom{E}&\phantom{E}&E\\
\hline
E&E&E&E&E&E&E\\
\hline
E&&&&E&&\\
\hline
\end{array}
&
\begin{array}{|c|c|c|c|c|c|c|}
\hline
\phantom{E}&E&E&\phantom{E}&\phantom{E}&\phantom{E}&\phantom{E}\\
\hline
&&&E&&&\\
\hline
&E&&&&&\\
\hline
\end{array}
\end{array}
$$
Each row with only one erasure is in
$\C^{(1)}_0$, so it can be corrected. After correcting the rows with
one erasure, since each of the four arrays is
in $\C^{(2)}_0$, and the first, second and fourth arrays,  contain
one row with two erasures while the remaining ones are erasure-free,
they are corrected without intervention of the other arrays.
Once these three arrays are corrected, the third array,
which has a row that is erasure-free, a row with two erasures and the
remaining row erased, is corrected
following the decoding algorithm of Theorem~\ref{theo2}.


By Theorems~\ref{cor00} and~\ref{lemma1}, code $\C'^{(3)}$
is an $[84,62,6]$ code, hence, it has the same rate as codes
$\C^{(2)}$ and $\C^{(3)}$ in Example~\ref{ex01}. However, both
$\C^{(2)}$ and $\C^{(3)}$ have minimum distance $d\eq 4$, while
$\C'^{(3)}$ has minimum distance $d\eq 6$.
}
\end{example}

\begin{example}
\label{ex02}
{\em This example is similar to the one given in~\cite{lxx}.
Consider the two nested 2-layer II codes 
$\C^{(2)}_0$ and $\C^{(2)}_1$ of Example~\ref{ex01}.
We construct two 3-layer II codes $\C^{(3)}_0$ and $\C^{(3)}_1$ with
$m_2\eq 2$ sharing the nested codes
$\{\uzero_{21}\}\eq\C^{(2)}_2\subset\C^{(2)}_1\subset\C^{(2)}_0$.
According to Definition~\ref{defmlayer}, for $\C^{(3)}_0$, let
$s_{0,0}\eq s_{0,1}\eq 1$ and $s_{0,2}\eq 0$ (hence,
$\C^{(3)}_0$ is a 3-layer 2-level II code), while for
$\C^{(3)}_1$, let $s_{1,0}\eq 0$, $s_{1,1}\eq 2$ and $s_{2,2}\eq 0$ (hence,
$\C^{(3)}_1$ is a 3-layer 1-level II code).
By Lemma~\ref{lemma2},
$\C_1^{(3)}\subset \C_0^{(3)}$.
According to Theorems~\ref{cor00} and~\ref{lemma1},
$\C^{(3)}_0$ is a $[42,32,4]$ 3-layer II code and $\C^{(3)}_1$ is a
$[42,30,4]$ 3-layer II
code. By Definition~\ref{ecap}, the erasure-correcting capability of
$\C^{(3)}_0$ is $((1,1,2),(1,2,3))$ while the one of
$\C^{(3)}_1$ is $((1,2,3),(1,2,3))$.

Next, using the nested codes
$\{\uzero_{42}\}\eq\C^{(3)}_2\subset\C^{(3)}_1\subset\C^{(3)}_0$, we
construct a 4-layer 2-level II code $\C^{(4)}_0$ with $m_3\eq 2$
and $s_1\eq s_2\eq 1$.
By~(\ref{uu}), the erasure-correcting capability of code $\C^{(4)}$
is $\uu^{(4)}\eq (((1,1,2),(1,2,3))\,,\,((1,2,3),(1,2,3)))$. Hence,
both the 4-layer II code $\C^{(4)}_0$ and the 3-layer II code
$\C^{(3)}$ of Example~\ref{ex01} can correct the same erasure
patterns, but code $\C^{(4)}_0$ has better
locality. In effect, if a row has three erasures and the remaining 11
rows are erasure-free, code $\C^{(3)}$ needs all 12 rows to recover
the erasures, while code $\C^{(4)}_0$ needs only 6 rows.

By Theorems~\ref{cor00} and~\ref{lemma1}, code $\C^{(4)}_0$, like
code $\C^{(3)}$ in Example~\ref{ex01}, is an $[84,62,4]$ code.
}
\end{example}

\begin{example}
\label{ex03}
{\em
We give another example of a 4-layer 2-level II code slightly different to
the one of Example~\ref{ex02}.

Assume that $n\eq 7$ and the 1-layer EII codes $\C^{(1)}_i$
over $GF(8)$ are as in Example~\ref{ex00}. Consider the two nested 2-layer
II codes $\C^{(2)}_1\subset\C^{(2)}_0$ of
Example~\ref{ex01}. In addition, take
two 2-layer 2-level II codes $\C^{(2)}_2$ and $\C^{(2)}_3$ with nested codes
$\C^{(1)}_2\subset\C^{(1)}_1\subset\C^{(1)}_0$ such that, for
$\C^{(2)}_2$, $s_0\eq 1$, $s_1\eq 2$ and $s_2\eq 0$, while for
$\C^{(2)}_3$, $s_0\eq 1$, $s_1\eq 0$ and $s_2\eq 2$. In
particular, notice that $\C^{(2)}_3\subset\C^{(2)}_2$, and, by
Theorems~\ref{cor00} and~\ref{lemma1},
$\C^{(2)}_2$ is a $[21,16,3]$ code and $\C^{(2)}_3$ is a $[21,14,4]$
code. By Definition~\ref{ecap}, the erasure-correcting capability of $\C^{(2)}_2$ is $(1,2,2)$
while the one of $\C^{(2)}_3$ is $(1,3,3)$.

Next, we construct two 3-layer II codes similarly to
Example~\ref{ex02} with $m_2\eq 2$. The first one, $\C^{(3)}_0$, is
the same as in Example~\ref{ex02}. We have seen that its erasure-correcting
capability is $((1,1,2),(1,2,3))$ and its minimum distance is 4.

The second one, $\C^{(3)}_2$, is a
2-level code with nested codes $\C^{(2)}_3\subset\C^{(2)}_2$, where
$\C^{(2)}_2$ and $\C^{(2)}_3$ were defined above, and $s_0\eq s_1\eq
1$. Hence
$\C_2^{(3)}\subset \C_0^{(3)}$. We have seen in Example~\ref{ex02}
that $\C{(3)}_0$ is a $[42,32,4]$ 3-layer II code with erasure-correcting
capability $((1,1,2),(1,2,3))$, while, by Theorems~\ref{cor00} and~\ref{lemma1},
$\C^{(3)}_2$ is a $[42,30,4]$ code.  By Definition~\ref{ecap}, its erasure-correcting
capability is $((1,2,2),(1,3,3))$.

Next, using the nested codes $\C^{(3)}_2\subset\C^{(3)}_0$, we
construct a 4-layer 2-level II code $\C^{(4)}_1$ with $m_3\eq 2$
and $s_1\eq s_2\eq 1$.

By Theorems~\ref{cor00} and~\ref{lemma1}, code $\C^{(4)}_1$ is an $[84,62,4]$ code.
By Definition~\ref{ecap}, its erasure-correcting capability is
$$\uu^{(4)}\eq (((1,1,2),(1,2,3)),((1,2,2),(1,3,3))).$$ We had seen
that the 4-layer II code $\C^{(4)}_0$ of Example~\ref{ex02} and the
3-layer II code
code $\C^{(3)}$ of Example~\ref{ex01} have the same
erasure-correcting capability. This is not true for $\C^{(4)}_1$
though. It has the same rate and minimum
distance as the previous two, but $\C^{(4)}_1$, can
correct two of the first three rows (i.e., rows 0, 1 and 2) with 3
erasures each
provided that the remaining ten rows are
erasure-free, and this is not true for $\C^{(3)}$ nor for
$\C^{(4)}_0$ (the same is true for consecutive rows 3, 4 and 5, 6, 7
and 8 and 9, 10 and 11).

For example, one such pattern correctable in $\C^{(4)}_1$ but not in
$\C^{(3)}$ nor in $\C^{(4)}_0$
consists of the following 4 arrays:

$$
\begin{array}{cccc}
\begin{array}{|c|c|c|c|c|c|c|}
\hline
\phantom{E}&E&\phantom{E}&E&\phantom{E}&E&\phantom{E}\\
\hline
E&&&&&&\\
\hline
&&E&E&&&E\\
\hline
\end{array}
&
\begin{array}{|c|c|c|c|c|c|c|}
\hline
\phantom{E}&\phantom{E}&E&\phantom{E}&\phantom{E}&E&\phantom{E}\\
\hline
&E&&&&E&\\
\hline
&&&&&&E\\
\hline
\end{array}
&
\begin{array}{|c|c|c|c|c|c|c|}
\hline
\phantom{E}&\phantom{E}&\phantom{E}&\phantom{E}&\phantom{E}&\phantom{E}&E\\
\hline
&&&E&&E&\\
\hline
E&&&&E&&E\\
\hline
\end{array}
&
\begin{array}{|c|c|c|c|c|c|c|}
\hline
\phantom{E}&E&E&\phantom{E}&\phantom{E}&\phantom{E}&\phantom{E}\\
\hline
&&&E&&&\\
\hline
&&&&&&E\\
\hline
\end{array}
\end{array}
$$

On the other hand, the erasure pattern on 4 arrays in
Example~\ref{ex01} is guaranteed to be correctable both in $\C^{(3)}$ and in
$\C^{(4)}_0$, but not in $\C^{(4)}_1$.
}
\end{example}

In this section, we have seen several examples of $[84,62]$ 2, 3 and
4-layer EII codes over $GF(16)$ and $GF(8)$. We will retake these examples in the
next section in which we give a general method for obtaining a
parity-check matrix of an $\ell$-layer EII code. The 4-layer II codes we
presented in Examples~\ref{ex02} and~\ref{ex03} have minimum distance
$d\eq 4$. However, it is possible to obtain an $[84,62]$ 4-layer II
code with minimum distace $d\eq 6$ as well. For example, we can construct,
using the same methods as in these two examples, a 4-layer II code with
erasure-correcting capability
$(((1,1,2),(1,2,3))\;,\;((1,1,2),(1,2,5)))$. By Theorem~\ref{lemma1},
this code has minimum distance 6, and it will be one of the codes
whose performance we will analize in Table~\ref{t1}.

We can increase the minimum distance of an $[84,62]$ 4-layer II
code even further. Take for example the 4-layer II code with
erasure-correcting capability
$(((0,0,1),(1,1,3))\;,\;((1,1,3),(2,3,6)))$. By Theorem~\ref{lemma1},
this code has minimum distance 7. However, with such a code, the
local erasure-correction on rows is lost.
Definition~\ref{defl1} allows to have $u_0\eq 0$, which would correspond to
code $\C_0$ being the whole space with minimum distance
$d_0\eq 1$ (no erasure-correcting capability). In effect, the 2-layer
(0,0,1) code can correct one erasure in one of three rows, while the
remaining two have to be erasure-free. For the
4-layer code with minimum distance 7, a correctable pattern must have
at least three consecutive rows with at most one erasure in them.

\section{Parity-check matrices of $\ell$-layer EII codes}
\label{secpc}
Given integers $m$ and $n$, let $\al$ be
an element in $GF(q)$ such that $\cO(\al)\geq\max\{m,n\}$.
Consider the following Vandermonde matrices of
rank $s$ for $s\leq \max\{m,n\}$ and $s\leq w\leq n$:

\begin{eqnarray}
\label{Hvws}
H^{(q)}_{s\,,\,w\,,\,v}&=&\left(
\begin{array}{ccccc}
1&\al^v&\al^{2v}&\ldots &\al^{(w-1)v}\\
1&\al^{v+1}&\al^{2(v+1)}&\ldots &\al^{(w-1)(v+1)}\\
\vdots&\vdots&\vdots&\ddots&\vdots\\
1&\al^{v+s-1}&\al^{2(v+s-1)}&\ldots &\al^{(w-1)(v+s-1)}\\
\end{array}
\right).
\end{eqnarray}

When the context is clear, we denote
$H^{(q)}_{s\,,\,w\,,\,v}$ simply as $H_{s\,,\,w\,,\,v}$. Also, we
denote by $I_n$ the $n\times n$ identity matrix, by
$\uzero_{m\times n}$ the $m\times n$ zero matrix, and by $A\otimes B$
the Kronecker product~\cite{ms} (also called the tensor product in
literature) of matrices $A$ and $B$.

The next theorem gives (recursively) a parity-check matrix for an EII
code according to Definition~\ref{defl1}.

\begin{theorem}
\label{l1}
{\rm
Let $\C$ be an EII code with nested codes $\C_i$ as given
by Definition~\ref{defl1}.
Let $\cH_0$ be the $u_0\times n$ parity-check matrix of $\C_0$ and,
assuming that the parity-check matrix of $\C_{i-1}$ is $\cH_{i-1}$
for $i\geq 1$, let

\begin{eqnarray}
\label{Hi}
\cH_i&=&
\left(\begin{array}{c}
\cH_{i-1}\\
\hline
B_i
\end{array}\right)
\end{eqnarray}
be the parity-check matrix of $\C_i$, where $B_i$ is an
$(u_i-u_{i-1})\times n$ matrix for $1\leq i\leq t-1$. Then a
parity-check matrix for code $\C$ is given by the
$(mu_0+n\hs_t+\sum_{i=1}^{t-1}(u_i-u_{i-1})\hs_i)\times n$ matrix

\begin{eqnarray}
\label{Hll}
\cH&=&\left(
\begin{array}{ccc}
I_{m}&\otimes &\cH_0\\
\hline
H_{\hs_{1}\,,\,m\,,\,0}&\otimes & B_1\\
\hline
H_{\hs_{2}\,,\,m\,,\,0}&\otimes & B_2\\
\hline
\vdots& \vdots& \vdots\\
\hline
H_{\hs_{t-1}\,,\,m\,,\,0}&\otimes & B_{t-1}\\
\hline
H_{\hs_{t}\,,\,m\,,\,0}&\otimes & I_n\\
\end{array}
\right)
\end{eqnarray}

}
\end{theorem}
\pf
We have to prove that $\uc\eq (\uc_0,\uc_1,\ldots,\uc_{m-1})\in\C$ if
and only if $\cH\uc^{\rm T}\eq\uzero_{w}$, with
$w\eq mu_0+n\hs_t+\sum_{i=1}^{t-1}(u_i-u_{i-1})\hs_i$.

Consider Definition~\ref{defl1}. Notice that $\uc_i\in\C_0$ for
$0\leq i\leq m-1$ if and only if
$\cH_0\uc_i^{\rm T}\eq\uzero_{u_0}$, if and only if
$(I_m\otimes\cH_0)\uc^{\rm T}\eq\uzero_{mu_0}$.

Next we prove that for every $i$, $1\leq i\leq t-1$,
$\bigoplus_{j=0}^{m-1}\al^{rj}\uc_j\in\C_i$ for
$0\leq r\leq \hs_{i}-1$ if and only if

\begin{eqnarray*}
(H_{\hs_{i}\,,\,m\,,\,0}\otimes
B_i)\uc^{\rm T}&\eq&\uzero_{(u_i-u_{i-1})\hs_{i}}.
\end{eqnarray*}

In effect, assume that for every $i$, $1\leq i\leq t-1$,
$\bigoplus_{j=0}^{m-1}\al^{rj}\uc_j\in\C_i$, where
$0\leq r\leq \hs_{i}-1$. We have to show that

\begin{eqnarray}
\label{Hri}
\bigoplus_{j=0}^{m-1}(\al^{rj}B_i)\uc_j^{\rm T}\;\;\eq\;\;
B_i\left(\bigoplus_{j=0}^{m-1}\al^{rj}\uc_j^{\rm T}\right)&=&\uzero_{u_i-u_{i-1}}.
\end{eqnarray}

Since $\bigoplus_{j=0}^{m-1}\al^{rj}\uc_j\in\C_i$ and,
since by~(\ref{Hi}) $B_i$ is part of the rows of the parity-check
matrix $\cH_i$ of $\C_i$, (\ref{Hri}) follows.

Conversely, assume that $(H_{\hs_{i}\,,\,m\,,\,0}\otimes
B_i)\uc^{\rm T}\eq\uzero_{(u_i-u_{i-1})\hs_{i}}$ for every $i$,
$1\leq i\leq t-1$. In particular, for every $r$, $0\leq r\leq
\hs_i-1$, $\left(\bigoplus_{j=0}^{m-1}\al^{rj}B_i\right)\uc^{\rm T}\eq\uzero_{u_i-u_{i-1}}$.
We have to show that 

\begin{eqnarray*}
\cH_i\left(\bigoplus_{j=0}^{m-1}\al^{rj}\uc_j\right)^{\rm
T}&=&\uzero_{u_i}\quad{\rm for}\quad 0\leq r\leq \hs_i-1,
\end{eqnarray*}
which will hold if and only if

\begin{eqnarray*}
\cH_0\left(\bigoplus_{j=0}^{m-1}\al^{rj}\uc_j\right)^{\rm
T}\eq\uzero_{u_0}\quad {\rm and}\quad
B_v\left(\bigoplus_{j=0}^{m-1}\al^{rj}\uc_j\right)^{\rm
T}\eq\uzero_{u_v-u_{v-1}}\quad{\rm for}\quad 1\leq v\leq i
\quad{\rm and}\quad 0\leq r\leq \hs_i-1.
\end{eqnarray*}

Notice that $B_v\left(\bigoplus_{j=0}^{m-1}\al^{rj}\uc_j\right)^{\rm
T}\eq\uzero_{u_v-u_{v-1}}$ for $0\leq r\leq \hs_v-1$, and since
$1\leq v\leq i$, $\hs_v\geq\hs_i$, in particular,

\begin{eqnarray*}
B_v\left(\bigoplus_{j=0}^{m-1}\al^{rj}\uc_j\right)^{\rm
T}&\eq&\uzero_{u_v-u_{v-1}}\quad {\rm for}\quad 0\leq r\leq\hs_i-1.
\end{eqnarray*}

Finally, notice that 
$\bigoplus_{j=0}^{m-1}\al^{jr}\uc_j\eq\uzero_n$ for $0\leq r\leq s_t-1$
if and only if $\left(H_{\hs_{t}\,,\,m\,,\,0}\otimes
I_n\right)\uc^{\rm T}\eq\uzero_{\hs_t n}$, completing the proof.\qed

Theorem~\ref{l1} allows us to obtain the parity-check matrix
of a 2-layer EII code in the next corollary.

\begin{corollary}
\label{ex0}
{\em
Let $\C^{(2)}$ be a 2-layer EII code  of length
$(m)(n)$ as given by
Definition~\ref{defmlayer}, where
$\{\uzero_n\}\eq\C^{(1)}_{t}\subset\C^{(1)}_{t-1}\subset\C^{(1)}_{t-2}\subset\cdots\subset
\C^{(1)}_{0}$ is the sequence of nested 1-level codes such that
$\C^{(1)}_i$ is an $[n,n-u_i,u_i+1]$ code with parity-check matrix
$H_{u_i\,,\,n\,,\,0}$ as given
by~(\ref{Hvws}), $0\leq u_0<u_1<\cdots <u_{t-1}<n$ and
$m\eq\sum_{i=0}^{t}s_i$, where $s_i\geq 0$ for $1\leq i\leq t$. Then,
the parity-check matrix of $\C^{(2)}$ is given by

\begin{eqnarray}
\label{Htl2}
\cH^{(2)}&=&\left(
\begin{array}{ccc}
I_{m}&\otimes & H_{u_0\,,\,n\,,\,0}\\
\hline
H_{\hs_{1}\,,\,m\,,\,0}&\otimes &  H_{u_1-u_{0}\,,\,n\,,\,u_{0}}\\
\hline
H_{\hs_{2}\,,\,m\,,\,0}&\otimes & H_{u_2-u_{1}\,,\,n\,,\,u_{1}}\\
\hline
\vdots&\vdots&\vdots \\
\hline
H_{\hs_{t-1}\,,\,m\,,\,0}&\otimes & H_{u_{t_1-1}-u_{t_1-2}\,,\,n\,,\,u_{t_1-2}}\\
\hline
H_{\hs_{t}\,,\,m\,,\,0}&\otimes &I_n
\end{array}
\right)
\end{eqnarray}
}
\end{corollary}

\noindent\pf
By~(\ref{Hvws}),

\begin{eqnarray}
\label{Huin0}
H_{u_i\,,\,n\,,\,0}&=&\left(
\begin{array}{c}
H_{u_{i-1}\,,\,n\,,\,0}\\
\hline
H_{u_i-u_{i-1}\,,\,n\,,\,u_{i-1}}\\
\end{array}
\right)
\end{eqnarray}

By~(\ref{Huin0}), taking $B_i\eq
H_{u_i-u_{i-1}\,,\,n\,,\,u_{i-1}}$ in~(\ref{Hi}) and~(\ref{Hll}), we obtain~(\ref{Htl2}).
\qed

It can be easily
proven that the parity-check matrix of a $t$-level EII code as given
by (25) in~\cite{b} is equivalent to $\cH^{(2)}$ as given
by~(\ref{Htl2}).

The next theorem 
gives a recursive construction for the parity-check
matrix of an $\ell$-layer EII code when $\ell\geq 3$.

\begin{theorem}
\label{Hrec}
{\em
Let $\ell\geq 3$ and consider an $\ell$-layer EII
code $\C^{(\ell)}$ of length $n_{\ell}\eq (m_{\ell-1})(m_{\ell-2})\ldots
(m_1)(n)$ as given by Definition~\ref{defmlayer}, where
$\{\uzero_{n_{\ell-1}}\}\eq\C^{(\ell-1)}_{t_{\ell-1}}\subset\C^{(\ell-1)}_{t_{\ell-1}-1}
\subset\C^{(\ell-1)}_{t_{\ell-1}-2}\subset\cdots
\subset\C^{(\ell-1)}_{0}$ is the sequence of $t_{\ell-1}+1$ nested
$(\ell-1)$-layer EII codes of length $n_{\ell-1}\eq (m_{\ell-2})(m_{\ell-3})\ldots
(m_1)(n)$ and $m_{\ell-1}\eq \sum_{i=0}^{t_{\ell-1}}s_i$, where $s_i\geq 0$ for $0\leq i\leq
t_{\ell-1}$. According to
Definition~\ref{defmlayer} and Corollary~\ref{ex0}, without loss of
generality, assume that the ($\ell-1$)-layer EII codes $\C^{(\ell-1)}_{i}$
share the same sequence of $(\ell-2)$-layer $[n_{\ell-2},n_{\ell-2}-u_j]$
nested EII codes
$\{\uzero_{n_{\ell-2}}\}\eq\C^{(\ell-2)}_{t_{\ell-2}}\subset
\C^{(\ell-2)}_{t_{\ell-2}-1}\subset\C^{(\ell-2)}_{t_{\ell-2}-2}\subset\cdots
\subset\C^{(\ell-2)}_{0}$, where $n_{\ell-2}\eq (m_{\ell-3})(m_{\ell-4})\ldots
(m_1)(n)$ for $\ell\geq 4$ and $n_1\eq n$ for $\ell\eq 3$. Denoting by $s_{i,j}$, $0\leq i\leq t_{\ell-1}$
and $0\leq j\leq t_{\ell-2}$, the $s_w$s of code $\C^{(\ell-1)}_i$ according to
Definition~\ref{defmlayer},
$\sum_{j=0}^{t_{\ell-2}}s_{i,j}\eq m_{\ell-2}$ and $s_{i,j}\geq
0$. Assuming that $B^{(1)}_j\eq H_{u_j-u_{j-1}\,,\,n\,,\,u_{j-1}}$
for $1\leq j\leq t_1-1$ when $\ell\eq 3$, let

\begin{eqnarray}
\label{HBiell}
B^{(\ell-1)}_i&=&\left(
\begin{array}{ccc}
H_{(\hs_{i,1}-\hs_{i-1,1})\,,\,m_{\ell-2}\,,\,\hs_{i-1,1}}&\otimes &  B^{(\ell-2)}_1\\
\hline
H_{(\hs_{i,2}-\hs_{i-1,2})\,,\,m_{\ell-2}\,,\,\hs_{i-1,2}}&\otimes & B^{(\ell-2)}_2\\
\hline
\vdots&\vdots&\vdots \\
\hline
H_{(\hs_{i,t_{\ell-2}}-\hs_{i-1,t_{\ell-2}-1})\,,\,m_{\ell-2}\,,\,\hs_{i-1,t_{\ell-2}-1}}&\otimes & B^{(\ell-2)}_{t_{\ell-2}-1}\\
\hline
H_{(\hs_{i,t_{\ell-2}}-\hs_{i-1,t_{\ell-2}})\,,\,m_{\ell-2}\,,\,\hs_{i-1,t_{\ell-2}}}&\otimes
& I_{n_{\ell-2}}
\end{array}
\right)
\end{eqnarray}
for $1\leq i\leq t_{\ell-1}-1$. Then, the parity-check matrix of the
$\ell$-layer $t_{\ell-1}$-level EII code $\C^{({\ell})}$
is given by

\begin{eqnarray}
\label{Hllell}
\cH^{(\ell )}&=&\left(
\begin{array}{ccc}
I_{m_{\ell-1}}&\otimes &\cH^{(\ell-1)}_0\\
\hline
H_{\hs_{1}\,,\,m_{\ell-1}\,,\,0}&\otimes & B^{(\ell-1)}_1\\
\hline
H_{\hs_{2}\,,\,m_{\ell-1}\,,\,0}&\otimes & B^{(\ell-1)}_2\\
\hline
\vdots& \vdots& \vdots\\
\hline
H_{\hs_{t_{\ell-1}-1}\,,\,m_{\ell-1}\,,\,0}&\otimes & B^{(\ell-1)}_{t_{\ell-1}-1}\\
\hline
H_{\hs_{t_{\ell-1}}\,,\,m_{\ell-1}\,,\,0}&\otimes & I_{n_{\ell-1}}
\end{array}
\right),
\end{eqnarray}
where $B^{(\ell-1)}_i$ is given by~(\ref{HBiell}). 
}
\end{theorem}

\noindent\pf
Let $\ell\geq 3$. By induction on $\ell$ and~(\ref{Hllell}), we
may assume that each
$(\ell-1)$-layer EII code
$\C^{(\ell-1)}_i$ is a code of length
$n_{\ell-1}\eq (m_{\ell-2})(m_{\ell-3})\ldots(m_1)(n)$ with parity-check matrix

\begin{eqnarray}
\label{Hllellm1}
\cH_i^{(\ell-1)}&=&\left(
\begin{array}{ccc}
I_{m_{\ell-2}}&\otimes &\cH^{(\ell-2)}_0\\
\hline
H_{\hs_{i,1}\,,\,m_{\ell-2}\,,\,0}&\otimes & B^{(\ell-2)}_1\\
\hline
H_{\hs_{i,2}\,,\,m_{\ell-2}\,,\,0}&\otimes & B^{(\ell-2)}_2\\
\hline
\vdots& \vdots& \vdots\\
\hline
H_{\hs_{i,t_{\ell-2}-1}\,,\,m_{\ell-2}\,,\,0}&\otimes & B^{(\ell-2)}_{t_{\ell-2}-1}\\
\hline
H_{\hs_{t_{\ell-2}}\,,\,m_{\ell-2}\,,\,0}&\otimes & I_{n_{\ell-2}}
\end{array}
\right),
\end{eqnarray}
if $\ell\geq 4$, where $B^{(\ell-2)}_i$ is given by taking $\ell-1$ instead of $\ell$
in~(\ref{HBiell}), while, when $\ell\eq 3$, $B^{(1)}_i\eq
H_{u_i-u_{i-1}\,,\,n\,,\,u_{i-1}}$ and, by~(\ref{Htl2}), 

\begin{eqnarray}
\label{Htl2bis}
\cH^{(2)}_i&=&\left(
\begin{array}{ccc}
I_{m_{1}}&\otimes & H_{u_0\,,\,n\,,\,0}\\
\hline
H_{\hs_{i,1}\,,\,m_{1}\,,\,0}&\otimes &  H_{u_1-u_{0}\,,\,n\,,\,u_{0}}\\
\hline
H_{\hs_{i,2}\,,\,m_{1}\,,\,0}&\otimes & H_{u_2-u_{1}\,,\,n\,,\,u_{1}}\\
\hline
\vdots&\vdots&\vdots \\
\hline
H_{\hs_{i,t_1-1}\,,\,m_{1}\,,\,0}&\otimes & H_{u_{t_1-1}-u_{t_1-2}\,,\,n\,,\,u_{t_1-2}}\\
\hline
H_{\hs_{i,t_1}\,,\,m_1\,,\,0}&\otimes &I_n
\end{array}
\right).
\end{eqnarray}

Since the codes $\C^{(\ell-1)}_i$ are nested, by Lemma~\ref{lemma2}, $\hs_{i,j}\geq
\hs_{i-1,j}$ for $1\leq i\leq t_{\ell-1}-1$ and $1\leq j\leq t_{\ell-2}$.
From~(\ref{Hi}) and~(\ref{Hllellm1}),
we can take $B_i\eq B^{(\ell-1)}_i$,
where $B^{(\ell-1)}_i$ is given by~(\ref{HBiell}),
so, 
by~(\ref{Hll}),
we obtain that the parity-check matrix of
$\C^{(\ell)}$ 
is given by~(\ref{Hllell}).
\qed

The next corollary simply applies Theorem~\ref{Hrec} to the case
$\ell\eq 3$. It is convenient to give it explicitly since it will
appear repeatedly in the examples.

\begin{corollary}
\label{ex1}
{\em
Consider a 3-layer EII code $\C^{(3)}$ of length
$(m_2)(m_1)(n)$ as given by
Definition~\ref{defmlayer}, where
$\{\uzero_{(m_1)(n)}\}\eq\C^{(2)}_{t_2}\subset\C^{(2)}_{t_2-1}\subset\C^{(2)}_{t_2-2}\subset\cdots
\subset\C^{(2)}_{0}$ is the sequence of $t_2+1$ nested
2-layer EII codes and $m_2\eq \sum_{i=0}^{t_2}s_i$, where $s_i\geq 0$ for
$0\leq i\leq t_2$.
According to
Definition~\ref{defmlayer} and Corollary~\ref{ex0}, without loss of
generality, we may assume that the 2-layer EII codes $\C^{(2)}_{i}$
share the same sequence of $t_1$ 1-layer $[n,n-u_j,u_j+1]$ nested EII codes
$\{\uzero_{n}\}\eq\C^{(1)}_{t_1}\subset\C^{(1)}_{t_1-1}\subset\C^{(1)}_{t_1-2}\subset\cdots
\subset\C^{(1)}_{0}$. Denoting by $s_{i,j}$, $0\leq i\leq t_2$
and $0\leq j\leq t_1$, the $s_w$s of code $\C^{(2)}_{i}$ according to
Definition~\ref{defmlayer},
$\sum_{j=0}^{t_1}s_{i,j}\eq m_1$ and $s_{i,j}\geq 0$. Let

\begin{eqnarray}
\label{HBi3}
B^{(2)}_i&=&\left(
\begin{array}{ccc}
H_{(\hs_{i,1}-\hs_{i-1,1})\,,\,m_{1}\,,\,\hs_{i-1,1}}&\otimes &  H_{u_1-u_{0}\,,\,n\,,\,u_{0}}\\
\hline
H_{(\hs_{i,2}-\hs_{i-1,2})\,,\,m_{1}\,,\,\hs_{i-1,2}}&\otimes & H_{u_2-u_{1}\,,\,n\,,\,u_{1}}\\
\hline
\vdots&\vdots&\vdots \\
\hline
H_{(\hs_{i,t_1-1}-\hs_{i-1,t_1-1})\,,\,m_{1}\,,\,\hs_{i-1,t_1-1}}&\otimes & H_{u_{t_1-1}-u_{t_1-2}\,,\,n\,,\,u_{t_1-2}}\\
\hline
H_{(\hs_{i,t_1}-\hs_{i-1,t_1})\,,\,m_1\,,\,\hs_{i-1,t_1}}&\otimes &I_n
\end{array}
\right)
\end{eqnarray}
for $1\leq i\leq t_2-1$. Then, the parity-check matrix of the
3-layer $t_2$-level EII code $\C^{(3)}$ obtained from the $t_2$ nested
2-layer EII codes $\C^{(2)}_j$ is given by

\begin{eqnarray}
\label{Hll3}
\cH^{(3)}&=&\left(
\begin{array}{ccc}
I_{m_2}&\otimes &\cH^{(2)}_0\\
\hline
H_{\hs_{1}\,,\,m_{2}\,,\,0}&\otimes & B^{(2)}_1\\
\hline
H_{\hs_{2}\,,\,m_{2}\,,\,0}&\otimes & B^{(2)}_2\\
\hline
\vdots& \vdots& \vdots\\
\hline
H_{\hs_{t_2-1}\,,\,m_{2}\,,\,0}&\otimes & B^{(2)}_{t_2-1}\\
\hline
H_{\hs_{t_2}\,,\,m_{2}\,,\,0}&\otimes & I_{(m_1)(n)}
\end{array},
\right)
\end{eqnarray}
where $B^{(2)}_i$ is given by~(\ref{HBi3}). \qed
}
\end{corollary}


%
%


Let us revisit next the examples of Section~\ref{sec3} to illustrate the
construction of parity-check matrices of $\ell$-layer EII codes.

\begin{example}
\label{ex00bis}
{\em
Consider the conditions of Example~\ref{ex00}. Both $\C^{(2)}_0$ and
$\C^{(2)}_1$ in Example~\ref{ex00}
have as nested codes the 1-layer
codes $\{\uzero_7\}\eq\C^{(1)}_2\subset\C^{(1)}_1\subset\C^{(1)}_0$
and we had $s_{0,0}\eq 5$, $s_{0,1}\eq 1$, $s_{0,2}\eq 0$, $s_{1,0}\eq 4$,
$s_{1,1}\eq 2$ and $s_{0,2}\eq 0$. Since $m_1\eq 6$, $u_0\eq 1$ and $u_1\eq 2$,
according to~(\ref{Htl2bis}), the parity-check matrix of $\C^{(2)}_0$
is the $7\times 42$ matrix

\begin{eqnarray*}
\cH^{(2)}_0&=&\left(
\begin{array}{ccc}
I_6&\otimes &H_{1\,,\,7\,,\,0}\\
\hline
H_{1,6,0}&\otimes &H_{1\,,\,7\,,\,1}
\end{array}
\right)\\
&=&
\left(
\begin{array}{cccccc}
H_{1\,,\,7\,,\,0}&\uzero_7&\uzero_7&\uzero_7&\uzero_7&\uzero_7\\
\uzero_7&H_{1\,,\,7\,,\,0}&\uzero_7&\uzero_7&\uzero_7&\uzero_7\\
\uzero_7&\uzero_7&H_{1\,,\,7\,,\,0}&\uzero_7&\uzero_7&\uzero_7\\
\uzero_7&\uzero_7&\uzero_7&H_{1\,,\,7\,,\,0}&\uzero_7&\uzero_7\\
\uzero_7&\uzero_7&\uzero_7&\uzero_7&H_{1\,,\,7\,,\,0}&\uzero_7\\
\uzero_7&\uzero_7&\uzero_7&\uzero_7&\uzero_7&H_{1\,,\,7\,,\,0}\\
\hline
H_{1\,,\,7\,,\,1}&H_{1\,,\,7\,,\,1}&H_{1\,,\,7\,,\,1}&H_{1\,,\,7\,,\,1}&H_{1\,,\,7\,,\,1}&H_{1\,,\,7\,,\,1}
\end{array}
\right)
\end{eqnarray*}

Similarly, according to~(\ref{Htl2bis}), the parity-check matrix of
$\C^{(2)}_1$ is the $8\times 42$ matrix

\begin{eqnarray*}
\cH^{(2)}_1&=&
\left(\begin{array}{ccc}
I_6&\otimes &H_{1\,,\,7\,,\,0}\\
\hline
H_{2,6,0}&\otimes &H_{1\,,\,7\,,\,1}
\end{array}
\right)\\
&=&
\left(
\begin{array}{cccccc}
H_{1\,,\,7\,,\,0}&\uzero_7&\uzero_7&\uzero_7&\uzero_7&\uzero_7\\
\uzero_7&H_{1\,,\,7\,,\,0}&\uzero_7&\uzero_7&\uzero_7&\uzero_7\\
\uzero_7&\uzero_7&H_{1\,,\,7\,,\,0}&\uzero_7&\uzero_7&\uzero_7\\
\uzero_7&\uzero_7&\uzero_7&H_{1\,,\,7\,,\,0}&\uzero_7&\uzero_7\\
\uzero_7&\uzero_7&\uzero_7&\uzero_7&H_{1\,,\,7\,,\,0}&\uzero_7\\
\uzero_7&\uzero_7&\uzero_7&\uzero_7&\uzero_7&H_{1\,,\,7\,,\,0}\\
\hline
H_{1\,,\,7\,,\,1}&H_{1\,,\,7\,,\,1}&H_{1\,,\,7\,,\,1}&H_{1\,,\,7\,,\,1}&H_{1\,,\,7\,,\,1}&H_{1\,,\,7\,,\,1}\\
H_{1\,,\,7\,,\,1}&\al H_{1\,,\,7\,,\,1}&\al^2H_{1\,,\,7\,,\,1}&\al^3H_{1\,,\,7\,,\,1}&\al^4H_{1\,,\,7\,,\,1}&\al^5H_{1\,,\,7\,,\,1}
\end{array}
\right)
\end{eqnarray*}

The 3-layer 2-level II code $\C^{(3)}$ of Example~\ref{ex00} has as nested codes
$\{\uzero_{42}\}\eq\C^{(2)}_2\subset\C^{(2)}_1\subset\C^{(2)}_0$,
$m_2\eq 2$ and $s_0\eq s_1\eq 1$ and $s_2\eq 0$.
Since $\hs_{1,1}\eq 2\geq \hs_{0,1}\eq 1$, according to~(\ref{HBi3}),

\begin{eqnarray*}
B^{(2)}_1&=&
H_{1\,,\,6\,,\,1}\otimes  H_{1\,,\,7\,,\,1}\\
&=&
\left(
\begin{array}{cccccc}
H_{1\,,\,7\,,\,1}&\al H_{1\,,\,7\,,\,1}&\al^2H_{1\,,\,7\,,\,1}&\al^3H_{1\,,\,7\,,\,1}&\al^4H_{1\,,\,7\,,\,1}&\al^5H_{1\,,\,7\,,\,1}
\end{array}
\right),
\end{eqnarray*}
so, according to~(\ref{Hll3}), the parity-check matrix of $\C^{(3)}$ is
the $15\times 84$ matrix

\begin{eqnarray*}
\cH^{(3)}&=&\left(
\begin{array}{ccc}
I_2&\otimes &\cH^{(2)}_0\\
\hline
H_{1,2,0}&\otimes &B^{(2)}_1
\end{array}
\right)\\
&=&
\left(
\begin{array}{cc}
\cH^{(2)}_0&\uzero_{7\times 42}\\
\uzero_{7\times 42}&\cH^{(2)}_0\\
\hline
B^{(2)}_1&B^{(2)}_1
\end{array}
\right)\\
&=&
\left(
\begin{array}{cc}
\left(\begin{array}{ccc}
I_6&\otimes &H_{1\,,\,7\,,\,0}\\
\hline
H_{1,6,0}&\otimes &H_{1\,,\,7\,,\,1}
\end{array}\right)&\uzero_{7\times 42}\\
\uzero_{7\times 42}&\left(\begin{array}{ccc}
I_6&\otimes &H_{1\,,\,7\,,\,0}\\
\hline
H_{1,6,0}&\otimes &H_{1\,,\,7\,,\,1}
\end{array}
\right)\\
\hline
H_{1\,,\,6\,,\,1}\otimes  H_{1\,,\,7\,,\,1}&H_{1\,,\,6\,,\,1}\otimes  H_{1\,,\,7\,,\,1}
\end{array}\right).
\end{eqnarray*}
}
\end{example}

\begin{example}
\label{ex01bis}
{\em
Let us take now the conditions of Example~\ref{ex01}.
Both $\C^{(2)}_0$ and
$\C^{(2)}_1$ in Example~\ref{ex01} have as nested codes the 1-layer
codes $\{\uzero_7\}\eq\C^{(1)}_3\subset\C^{(1)}_2\subset\C^{(1)}_1\subset\C^{(1)}_0$, 
and $u_0\eq 1$, $u_1\eq 2$, $u_2\eq 2$,
$m_1\eq 3$,
$s_{0,0}\eq 2$, $s_{0,1}\eq 1$, $s_{0,2}\eq 0$, $s_{0,3}\eq 0$,
$s_{1,0}\eq s_{1,1}\eq s_{1,2}\eq 1$ and $s_{1,3}\eq 0$.
According to~(\ref{Htl2bis}), the parity-check matrix of the 2-layer
2-level code $\C^{(2)}_0$
is the $4\times 21$ matrix

\begin{eqnarray}
\nonumber
\cH^{(2)}_0&=&\left(
\begin{array}{ccc}
I_3&\otimes &H_{1\,,\,7\,,\,0}\\
\hline
H_{1,3,0}&\otimes &H_{1\,,\,7\,,\,1}
\end{array}
\right)\\
\label{H02}
&=&
\left(
\begin{array}{ccc}
H_{1\,,\,7\,,\,0}&\uzero_7&\uzero_7\\
\uzero_7&H_{1\,,\,7\,,\,0}&\uzero_7\\
\uzero_7&\uzero_7&H_{1\,,\,7\,,\,0}\\
\hline
H_{1\,,\,7\,,\,1}&H_{1\,,\,7\,,\,1}&H_{1\,,\,7\,,\,1}
\end{array}
\right). 
\end{eqnarray}

Similarly, by~(\ref{Htl2bis}), the parity-check matrix of the 2-layer
3-level code $\C^{(2)}_1$
is the $6\times 21$ matrix

\begin{eqnarray}
\nonumber
\cH^{(2)}_1&=&\left(
\begin{array}{ccc}
I_3&\otimes &H_{1,7,0}\\
\hline
H_{2,3,0}&\otimes &H_{1,7,1}\\
\hline
H_{1,3,0}&\otimes &H_{1,7,2}
\end{array}
\right)\\
\label{H12}
&=&
\left(
\begin{array}{ccc}
H_{1,7,0}&\uzero_7&\uzero_7\\
\uzero_7&H_{1,7,0}&\uzero_7\\
\uzero_7&\uzero_7&H_{1,7,0}\\
\hline
H_{1,7,1}&H_{1,7,1}&H_{1,7,1}\\
H_{1,7,1}&\al H_{1,7,1}&\al^2H_{1,7,1}\\
\hline
H_{1,7,2}&H_{1,7,2}&H_{1,7,2}\\
\end{array}
\right). 
\end{eqnarray}

The 3-layer 2-level EII code $\C^{(3)}$ of Example~\ref{ex01} has as nested codes
$\{\uzero_{21}\}\eq\C^{(2)}_2\subset\C^{(2)}_1\subset\C^{(2)}_0$,
$t_2\eq 2$, $m_2\eq 4$, $s_0\eq 1$, $s_1\eq 3$ and $s_2\eq 0$.
Since $\hs_{0,1}\eq 1$, $\hs_{0,2}\eq 0$, $\hs_{0,3}\eq 0$,
$\hs_{1,1}\eq 2$, $\hs_{1,2}\eq 1$ and $\hs_{1,3}\eq 0$,
according to~(\ref{HBi3}),

\begin{eqnarray}
\nonumber
B^{(2)}_1&=&
\left(
\begin{array}{ccc}
H_{1,3,1}&\otimes&H_{1,7,1}\\
H_{1,3,0}&\otimes&H_{1,7,2}\\
\end{array}
\right)\\
\label{B21}
&=&
\left(
\begin{array}{ccc}
H_{1,7,1}&\al H_{1,7,1}&\al^2H_{1,7,1}\\
H_{1,7,2}&H_{1,7,2}&H_{1,7,2}
\end{array}\right),
\end{eqnarray}
so, according to~(\ref{Hll3}), the parity-check matrix of $\C^{(3)}$ is
the $22\times 84$ matrix

\begin{eqnarray*}
\cH^{(3)}&=&\left(
\begin{array}{ccc}
I_4&\otimes &\cH^{(2)}_0\\
\hline
H_{3,4,0}&\otimes &B^{(2)}_1
\end{array}
\right)\\
&=&
\left(
\begin{array}{cccc}
\cH^{(2)}_0&\uzero_{4\times 7}&\uzero_{4\times 7}&\uzero_{4\times 7}\\
\uzero_{4\times 7}&\cH^{(2)}_0&\uzero_{4\times 7}&\uzero_{4\times 7}\\
\uzero_{4\times 7}&\uzero_{4\times 7}&\cH^{(2)}_0&\uzero_{4\times 7}\\
\uzero_{4\times 7}&\uzero_{4\times 7}&\uzero_{4\times 7}&\cH^{(2)}_0\\
\hline
B^{(2)}_1&B^{(2)}_1&B^{(2)}_1&B^{(2)}_1\\
B^{(2)}_1&\al B^{(2)}_1&\al^2B^{(2)}_1&\al^3B^{(2)}_1\\
B^{(2)}_1&\al^2 B^{(2)}_1&\al^4B^{(2)}_1&\al^6B^{(2)}_1\\
\end{array}
\right),
\end{eqnarray*}
where $\cH^{(2)}_0$ is given by~(\ref{H02}) and $B^{(2)}_1$ by~(\ref{B21}).

Consider next the 2-layer
2-level II code $\C^{(2)}$ over $GF(16)$ in Example~\ref{ex01}, then, according
to~(\ref{Hll}), its parity-check matrix is given by
the $22\times 84$ matrix

\begin{eqnarray*}
\cH^{(2)}&=&\left(
\begin{array}{ccc}
I_{12}&\otimes &H_{1,7,0}\\
\hline
H_{7,12,0}&\otimes &H_{1,7,1}\\
\hline
H_{3,12,0}&\otimes &H_{1,7,2}\\
\end{array}
\right).
\end{eqnarray*}
}
\end{example}

\begin{example}
\label{ex04bis}
{\em
Consider the conditions of Example~\ref{ex04}.
Since $m_1\eq 3$,
$s_{0,0}\eq 2$, $s_{0,1}\eq 1$, $s_{0,2}\eq 0$ and $s_{1,0}\eq s_{1,1}\eq s_{1,2}\eq 1$.
The parity-check matrix $\cH^{(2)}_0$ of the 2-layer
2-level II code $\C^{(2)}_0$ is
given by~(\ref{H02}), while, by~(\ref{Htl2bis}),
the parity-check matrix of $\C'^{(2)}_1$
is the $12\times 21$ matrix

\begin{eqnarray}
\nonumber
\cH'^{(2)}_1&=&\left(
\begin{array}{ccc}
I_3&\otimes &H_{1,7,0}\\
\hline
H_{2,3,0}&\otimes &H_{1,7,1}\\
\hline
H_{1,3,0}&\otimes &I_7
\end{array}
\right)\\
\label{H'12}
&=&
\left(
\begin{array}{ccc}
H_{1,7,0}&\uzero_7&\uzero_7\\
\uzero_7&H_{1,7,0}&\uzero_7\\
\uzero_7&\uzero_7&H_{1,7,0}\\
\hline
H_{1,7,1}&H_{1,7,1}&H_{1,7,1}\\
H_{1,7,1}&\al H_{1,7,1}&\al^2H_{1,7,1}\\
\hline
I_7&I_7&I_7
\end{array}
\right).
\end{eqnarray}

Since $\C'^{(2)}_1$ is a [21,11] code, the number of parities is 10,
so two of the rows of $\cH'^{(2)}_1$ as given by~(\ref{H'12}) are
dependent since the matrix
has rank 10. We can delete the last two rows of $\cH'^{(2)}_1$
to obtain a parity-check matrix of rank 10.

The 3-layer 2-level EII code $\C'^{(3)}$ of Example~\ref{ex01} has as nested codes
$\{\uzero_{21}\}\eq\C^{(2)}_2\subset\C'^{(2)}_1\subset\C^{(2)}_0$, $m_2\eq 4$, $s_0\eq 3$ and $s_1\eq 1$.
Since $\hs_{1,1}\eq 2\geq \hs_{0,1}\eq 1$ and $\hs_{1,2}\eq 1\geq \hs_{0,2}\eq 0$, according to~(\ref{HBi3}),

\begin{eqnarray}
\nonumber
B'^{(2)}_1&=&\left(
\begin{array}{ccc}
H_{1,3,1}&\otimes &H_{1,7,1}\\
H_{1,3,0}&\otimes &I_7
\end{array}
\right)\\
\label{B'21}
&=&
\left(
\begin{array}{ccc}
H_{1,7,1}&\al H_{1,7,1}&\al^2H_{1,7,1}\\
I_7&I_7&I_7\\
\end{array}
\right),
\end{eqnarray}
so, according to~(\ref{Hll3}), the parity-check matrix of the 3-layer
2-level II code $\C'^{(3)}$ of Example~\ref{ex04} is
the $24\times 84$ matrix

\begin{eqnarray*}
\cH'^{(3)}&=&\left(
\begin{array}{ccc}
I_4&\otimes &\cH^{(2)}_0\\
\hline
H_{1,4,0}&\otimes &B'^{(2)}_1\\
\end{array}
\right)\\
&=&
\left(
\begin{array}{cccc}
\cH^{(2)}_0&\uzero_{4\times 7}&\uzero_{4\times 7}&\uzero_{4\times 7}\\
\uzero_{4\times 7}&\cH^{(2)}_0&\uzero_{4\times 7}&\uzero_{4\times 7}\\
\uzero_{4\times 7}&\uzero_{4\times 7}&\cH^{(2)}_0&\uzero_{4\times 7}\\
\uzero_{4\times 7}&\uzero_{4\times 7}&\uzero_{4\times 7}&\cH^{(2)}_0\\
\hline
B'^{(2)}_1&B'^{(2)}_1&B'^{(2)}_1&B'^{(2)}_1\\
\end{array}
\right),
\end{eqnarray*}
where $B'^{(2)}_1$ is given by~(\ref{B'21}).

Matrix $\cH'^{(3)}$ has rank 22 since $\C'^{(3)}$ has dimension 22.
We can delete the last two rows of $\cH'^{(3)}$ to obtain a
$22\times 84$ parity-check matrix of $\C'^{(3)}$.
}
\end{example}

\begin{example}
\label{ex02bis}
{\em
We revisit now Example~\ref{ex02}.
We had 
the 3-layer nested codes
$\{\uzero_{42}\}\eq\C^{(3)}_2\subset\C^{(3)}_1\subset \C^{(3)}_0$ sharing the 2-layer nested codes
$\{\uzero_{21}\}\eq\C^{(2)}_2\subset\C^{(2)}_1\subset\C^{(2)}_0$,
where $s_{0,0}\eq s_{0,1}\eq 1$ and $s_{0,2}\eq 0$ correspond to
$\C^{(3)}_0$ and $s_{1,0}\eq 0$, $s_{1,1}\eq 2$ and $s_{1,2}\eq 0$
correspond to $\C^{(3)}_1$.

Proceeding as in previous examples, applying Corollary~\ref{ex1}, we
can verify that the parity-check
matrix of $\C^{(3)}_0$ is given by the $10\times 42$ matrix

\begin{eqnarray}
\nonumber
\cH^{(3)}_0&=&\left(
\begin{array}{ccc}
I_2&\otimes &\cH^{(2)}_0\\
\hline
H_{1,2,0}&\otimes &B^{(2)}_1\\
\end{array}
\right)\\
\label{eqH30}
&=&
\left(
\begin{array}{cc}
\cH^{(2)}_0&\uzero_{4\times 21}\\
\uzero_{4\times 21}&\cH^{(2)}_0\\
\hline
B^{(2)}_1&B^{(2)}_1
\end{array}
\right),
\end{eqnarray}
where $B^{(2)}_1$ is given by~(\ref{B21}).

Similarly, 
the parity-check
matrix of $\C^{(3)}_1$ is given by the $12\times 42$ matrix

\begin{eqnarray*}
\nonumber
\cH^{(3)}_1&=&\left(
\begin{array}{ccc}
I_2&\otimes &\cH^{(2)}_0\\
\hline
H_{2,2,0}&\otimes &B^{(2)}_1
\end{array}
\right)\\
\label{eqHp31}
&=&
\left(
\begin{array}{cc}
\cH^{(2)}_0&\uzero_{4\times 21}\\
\uzero_{4\times 21}&\cH^{(2)}_0\\
\hline
B^{(2)}_1&B^{(2)}_1\\
B^{(2)}_1&\al B^{(2)}_1
\end{array}
\right)
\end{eqnarray*}

Consider the 4-layer 2-level II code $\C^{(4)}_0$ of
Example~\ref{ex02} with nested codes $\{\uzero_{42}\}\eq\C^{(3)}_2\subset\C^{(3)}_1\subset \C^{(3)}_0$.
We had, $s_0\eq s_1\eq 1$, $s_2\eq 0$, $m_3\eq m_2\eq 2$, $s_{0,0}\eq s_{0,1}\eq
1$, $s_{0,2}\eq 0$, $s_{1,0}\eq 0$, $s_{1,1}\eq 2$ and $s_{1,2}\eq
0$. Since $\hs_{1,1}\eq 2$ and $\hs_{0,1}\eq 1$, according to~(\ref{HBiell}),

\begin{eqnarray}
\label{B31}
B^{(3)}_1&=&\left(
\begin{array}{cc}
B^{(2)}_1&\al B^{(2)}_1
\end{array}
\right),
\end{eqnarray}
where $B^{(2)}_1$ is given by~(\ref{B21}), while
according to~(\ref{Hll}), (\ref{eqH30}) and~(\ref{B31}), the parity-check matrix of
$\C^{(4)}_0$ is given by the $22\times 84$ matrix

\begin{eqnarray*}
\cH^{(4)}_0&=&\left(
\begin{array}{ccc}
I_{2}&\otimes &\cH^{(3)}_0\\
\hline
H_{1,2,0}&\otimes &B^{(3)}_1\\
\end{array}
\right)\\
&=&
\left(
\begin{array}{cc}
\cH^{(3)}_0&\uzero_{10\times 42}\\
\uzero_{10\times 42}&\cH^{(3)}_0\\
\hline
B^{(3)}_1&B^{(3)}_1\\
\end{array}
\right).
\end{eqnarray*}
}
\end{example}

\begin{example}
\label{ex5}
{\em
Assume that we take the four nested codes 1-layer II codes
$\{\uzero_7\}\eq\C^{(1)}_3\subset\C^{(1)}_2\subset\C^{(1)}_1\subset\C^{(1)}_0$, where $\C^{(1)}_i$
is a $[7,7-i-1,i+1]$ code over $GF(8)$. With these three codes, we
construct, using Definition~\ref{defmlayer}, the three nested 2-layer
II codes
$\C^{(2)}_2\subset\C^{(2)}_1\subset\C^{(2)}_0$  with
$m_1\eq 5$ and $n\eq 7$, where
$s_{0,0}\eq 4$, $s_{0,1}\eq 1$, $s_{0,2}\eq 0$, $s_{0,3}\eq 0$,
$s_{1,0}\eq 3$, $s_{1,1}\eq 2$, $s_{1,2}\eq 0$, $s_{1,3}\eq 0$,
$s_{2,0}\eq 3$, $s_{2,1}\eq 1$, $s_{2,2}\eq 1$, $s_{2,3}\eq 0$, (hence,
$s_{i,0}+s_{i,1}+s_{i,2}\eq 5$ for $0\leq i\leq 2$), let
$\cH^{(2)}_i$ be the parity-check matrix of code $\C^{(2)}_i$
according to~(\ref{Htl2bis}) and assume that
we want to construct the parity-check matrix $\cH^{(3)}_0$ of a
3-layer 2-level II code $\C^{(3)}_0$ with
$m_2\eq 4$,  $s_0\eq 2$, $s_1\eq 2$  and $s_1\eq 0$ (hence,
$s_0+s_1\eq 4$). Notice that the erasure-correcting capability of
$\C^{(2)}_0$ is $(1,1,1,1,2)$, the one of $\C^{(2)}_1$ is
$(1,1,1,2,2)$ and the one of $\C^{(2)}_2$ is $(1,1,1,2,3)$. Hence,
the erasure-correcting capability of
$\C^{(3)}_0$ is
$\left(\;(1,1,1,1,2),(1,1,1,1,2),(1,1,1,2,2),(1,1,1,2,2)\;\right)$
and its minimum distance, according to Theorem~\ref{lemma1}, is
$d^{(3)}_0\eq 3$.

By~(\ref{HBi3}),

\begin{eqnarray}
\label{BB21}
B^{(2)}_1&=&\left(
\begin{array}{ccc}
H_{1\,,\,5\,,\,1}&\otimes &  H_{1\,,\,7\,,\,1}\\
\end{array}
\right).
\end{eqnarray}

Applying~(\ref{Htl2bis}), 
(\ref{Hll3}) and~(\ref{BB21}), we obtain that the parity-check matrix of
$\C^{(3)}_0$ is the $26\times 140$ matrix

\begin{eqnarray}
\nonumber
\cH^{(3)}_0&=&
\left(
\begin{array}{ccc}
I_{4}&\otimes &\cH^{(2)}_0\\
\hline
H_{2\,,\,4\,,\,0}&\otimes & B^{(2)}_1
\end{array}\right)\\
\label{H30}
&=&
\left(
\begin{array}{ccc}
I_{4}&\otimes &\left(
\begin{array}{ccc}
I_{5}&\otimes & H_{1\,,\,7\,,\,0}\\
\hline
H_{1\,,\,5\,,\,0}&\otimes &  H_{1\,,\,7\,,\,1}\\
\end{array}
\right)\\
\hline
H_{2\,,\,4\,,\,0}&\otimes & \left(
\begin{array}{ccc}
H_{1\,,\,5\,,\,1}&\otimes &  H_{1\,,\,7\,,\,1}\\
\end{array}
\right)\\
\end{array}
\right).
\end{eqnarray}

Consider next the parity-check matrix $\cH^{(3)}_1$ of a
3-layer 3-level II code $\C^{(3)}_1$ with
$m_2\eq 4$,  $s_0\eq 2$, $s_1\eq 1$ and $s_2\eq 1$. Notice that the
erasure-correcting capability of $\C^{(3)}_1$ is
$\left(\;(1,1,1,1,2),(1,1,1,1,2),(1,1,1,2,2),(1,1,1,2,3)\;\right)$
and its minimum distance, according to Theorem~\ref{lemma1}, is
$d^{(3)}_1\eq 4$.

By~(\ref{HBi3}),

\begin{eqnarray}
\label{BB22}
B^{(2)}_2&=&\left(
\begin{array}{ccc}
H_{(1\,,\,5\,,\,0}&\otimes &  H_{1\,,\,7\,,\,2}\\
\end{array}
\right)
\end{eqnarray}

Applying~(\ref{Htl2bis}), (\ref{BB21}), (\ref{BB22}) and~(\ref{Hll3}), we obtain the $27\times 140$ matrix

\begin{eqnarray}
\nonumber
\cH^{(3)}_1&=&
\left(
\begin{array}{ccc}
I_{4}&\otimes &\cH^{(2)}_0\\
\hline
H_{2\,,\,4\,,\,0}&\otimes & B^{(2)}_1\\
\hline
H_{1\,,\,4\,,\,0}&\otimes & B^{(2)}_2
\end{array}\right)\\
\label{H31}
&=&
\left(
\begin{array}{ccc}
I_{4}&\otimes &\left(
\begin{array}{ccc}
I_{5}&\otimes & H_{1\,,\,7\,,\,0}\\
\hline
H_{1\,,\,5\,,\,0}&\otimes &  H_{1\,,\,7\,,\,1}\\
\end{array}
\right)\\
\hline
H_{2\,,\,4\,,\,0}&\otimes & \left(
\begin{array}{ccc}
H_{1\,,\,5\,,\,1}&\otimes &  H_{1\,,\,7\,,\,1}\\
\hline
\end{array}
\right)\\
\hline
H_{1\,,\,4\,,\,0}&\otimes & \left(
\begin{array}{ccc}
H_{1\,,\,5\,,\,0}&\otimes & H_{1\,,\,7\,,\,2}\\
\end{array}
\right)\\
\end{array}
\right).
\end{eqnarray}

Finally, consider the parity-check matrix $\cH^{(3)}_2$ of a
3-layer 3-level code $\C^{(3)}_2$ with
$m_2\eq 4$,  $s_0\eq 1$, $s_1\eq 2$ and $s_2\eq 1$.
The
erasure-correcting capability of $\C^{(3)}_2$ is
$((1,1,1,1,2),(1,1,1,2,2),(1,1,1,2,2),(1,1,1,2,3))$ and its minimum distance, according to Theorem~\ref{lemma1}, is
$d^{(3)}_2\eq 4$.

Applying~(\ref{Htl2bis}) (\ref{BB21}), (\ref{BB22}) and~(\ref{Hll3}), we obtain the $28\times 140$ matrix

\begin{eqnarray}
\nonumber
\cH^{(3)}_2&=&
\left(
\begin{array}{ccc}
I_{4}&\otimes &\cH^{(2)}_0\\
\hline
H_{3\,,\,4\,,\,0}&\otimes & B^{(2)}_1\\
\hline
H_{1\,,\,4\,,\,0}&\otimes & B^{(2)}_2
\end{array}\right)\\
\label{H32}
&=&
\left(
\begin{array}{ccc}
I_{4}&\otimes &\left(
\begin{array}{ccc}
I_{5}&\otimes & H_{1\,,\,7\,,\,0}\\
\hline
H_{1\,,\,5\,,\,0}&\otimes &  H_{1\,,\,7\,,\,1}\\
\end{array}
\right)\\
\hline
H_{3\,,\,4\,,\,0}&\otimes & \left(
\begin{array}{ccc}
H_{1\,,\,5\,,\,1}&\otimes &  H_{1\,,\,7\,,\,1}\\
\hline
\end{array}
\right)\\
\hline
H_{1\,,\,4\,,\,0}&\otimes & \left(
\begin{array}{ccc}
H_{1\,,\,5\,,\,0}&\otimes & H_{1\,,\,7\,,\,2}\\
\end{array}
\right)\\
\end{array}
\right).
\end{eqnarray}

We will use the three nested codes
$\C^{(3)}_2\subset\C^{(3)}_1\subset\C^{(3)}_0$ to construct the
parity-check matrix of a 4-layer II code in the next example.
}
\end{example}

\begin{example}
\label{ex4}
{\rm
Consider the four nested 3-layer II codes
$\{\uzero_{140}\}\eq\C^{(3)}_3\subset\C^{(3)}_2\subset\C^{(3)}_1\subset\C^{(3)}_0$ of Example~\ref{ex5}.
Assume that we want to construct the parity-check matrix $\cH^{(4)}$ of a
4-layer 3-level II code $\C^{(4)}$ with
$m_3\eq 3$ and  $s_0\eq s_1\eq s_2\eq 1$ using~(\ref{Hllell})
and~(\ref{HBiell}) in Theorem~\ref{Hrec}.

Explicitly, since $s_{0,0}\eq 2$, $s_{0,1}\eq 2$, $s_{0,2}\eq 0$, $s_{0,3}\eq 0$,
$s_{1,0}\eq 2$, $s_{1,1}\eq 1$, $s_{1,2}\eq 1$, $s_{1,3}\eq 0$,
$s_{2,0}\eq 1$, $s_{2,1}\eq 2$, $s_{2,2}\eq 1$ and $s_{2,3}\eq 0$,
according to~(\ref{HBiell}), (\ref{BB21}) and (\ref{BB22}),

\begin{eqnarray}
\nonumber
B^{(3)}_1&=&
\begin{array}{ccc}
H_{1\,,\,4\,,\,0}&\otimes &B^{(2)}_2
\end{array}\\
\label{B1bis}
&=&
\begin{array}{ccc}
H_{1\,,\,4\,,\,0}&\otimes &
\left(
\begin{array}{ccc}
H_{1\,,\,5\,,\,0}&\otimes &H_{1\,,\,7\,,\,2}
\end{array}
\right)
\end{array}
\end{eqnarray}
and
\begin{eqnarray}
\nonumber
B^{(3)}_2&=&
\begin{array}{ccc}
H_{1\,,\,4\,,\,2}&\otimes &B^{(2)}_1
\end{array}\\
\label{B2bis}
&=&
\begin{array}{ccc}
H_{1\,,\,4\,,\,2}&\otimes &
\left(
\begin{array}{ccc}
H_{1\,,\,5\,,\,1}&\otimes &H_{1\,,\,7\,,\,1}
\end{array}
\right)
\end{array}.
\end{eqnarray}
According to~(\ref{Hll}),
(\ref{Hll3}), (\ref{H30}), (\ref{B1bis}) and~(\ref{B2bis}), the
parity-check matrix of $\C^{(4)}$ is given by the $81\times 420$
matrix

\begin{eqnarray}
\nonumber
\cH^{(4)}&=&
\left(
\begin{array}{ccc}
I_{3}&\otimes &\cH^{(3)}_0\\
\hline
H_{2\,,\,3\,,\,0}&\otimes &B^{(3)}_1\\
\hline
H_{1\,,\,3\,,\,0}&\otimes &B^{(3)}_2\\
\end{array}
\right)\\
\label{H4}
&=&
\left(
\begin{array}{ccc}
I_{3}&\otimes &
\left(
\begin{array}{ccc}
I_{4}&\otimes &\left(
\begin{array}{ccc}
I_{5}&\otimes & H_{1\,,\,7\,,\,0}\\
\hline
H_{1\,,\,5\,,\,0}&\otimes &  H_{1\,,\,7\,,\,1}\\
\end{array}
\right)\\
\hline
H_{2\,,\,4\,,\,0}&\otimes & \left(
\begin{array}{ccc}
H_{1\,,\,5\,,\,1}&\otimes &  H_{1\,,\,7\,,\,1}\\
\end{array}
\right)\\
\end{array}
\right)\\
\hline
H_{2\,,\,3\,,\,0}&\otimes &
\left(\begin{array}{ccc}
H_{1\,,\,4\,,\,0}&\otimes &
\left(
\begin{array}{ccc}
H_{1\,,\,5\,,\,0}&\otimes &H_{1\,,\,7\,,\,2}
\end{array}
\right)
\end{array}\right)\\
\hline
H_{1\,,\,3\,,\,0}&\otimes &\left(\begin{array}{ccc}
H_{1\,,\,4\,,\,2}&\otimes &
\left(
\begin{array}{ccc}
H_{1\,,\,5\,,\,1}&\otimes &H_{1\,,\,7\,,\,1}
\end{array}
\right)
\end{array}\right)\\
\end{array}
\right).
\end{eqnarray}
}
\end{example}

The parity-check of a code allows for decoding erasures in a
traditional way, that is, by inverting the submatrix with columns
corresponding to the erasures. The decoding algorithm for erasures as
illustrated in Theorem~\ref{theo2} is certainly more efficient than
the straightforward inverting method, but the advantage of using the
parity-check matrix is that some extra erasures may be corrected. The
decoding algorithm in Theorem~\ref{theo2} can only correct those
erasures that can be guaranteed to be corrected according to
Theorem~\ref{theo2}, but there may be more possible correctable
erasures. This point has also been made in other papers~\cite{bh2,b}.
We will elaborate further in Section~\ref{secperf}, in particular, in
Table~\ref{t1}.

\section{Average number of uncorrectable erasures}
\label{secperf}
In this section, as done in~\cite{bh} for 2-layer II codes, we examine the average
number of erasures causing an uncorrectable pattern in $\ell$-layer
EII codes. Let us call this
parameter the Average Number of Erasures to Failure (ANETF). The ANETF
is more relevant than the minimum distance of the code when failures
do not occur all at the same time, but one after the other, for
example, by arriving following a Poisson distribution~\cite{bgm,gm,gm2}.

So, assume that failures (erasures) occur consecutively, one after
the other. The question is, given an $\ell$-layer EII code, what is
its ANETF? In~\cite{bh}, it was found that in some
cases, 2-layer II codes having lower
minimum distance than others, nevertheless had higher ANETF.

As an example, we take a number of multiple layer EII codes with rate
62/84, as illustrated in Table~\ref{t1}. We choose this rate
since~\cite{z3} gives
the example of a 3-layer 2-level II code with erasure-correcting capability
$\left(\,(1,1,2),(1,2,3),(1,2,3),(1,2,3)\,\right)$, while~\cite{lxx}
gives a 4-layer 2-level II code with
erasure-correcting capability
$\left(\,((1,1,2),(1,2,3)))\,,\,((1,2,3),(1,2,3))\,\right)$, both
cases with rate 62/84. We
retake these two examples in
Table~\ref{t1}, as well as several others with the same rate.

The first column in Table~\ref{t1} gives the $\ell$ corresponding to
the layer of the EII code
described, in this case, $1\leq\ell\leq 4$. The second column gives
the erasure-correcting capability of the code. The third column gives
the minimum distance of the code according to Theorem~\ref{lemma1}.
The fourth column gives the result of simulations where the ANETF was
computed in two ways. One was by using the algorithm of
Theorem~\ref{theo2}, where,
each time the erasure-correcting capability in the third column is
exceeded, an uncorrectable pattern is declared. The second
computation of the ANETF in
the fourth column is in parenthesis and with an asterisk. It is
obtained by doing erasure-decoding using the parity-check of the code
as given in Section~\ref{secpc} (certainly, both decoding methods can
be combined: if an erasure pattern exceeds the erasure-correcting
capability of the code, decoding may be attempted using the
parity-check matrix). We can
see that the ANETF improves considerably when using the
parity-check matrix to correct erasure patterns that exceed the
erasure-correcting capability of the codes. In some cases, the
improvement is dramatic. For example, for the 4-layer code in the
next to last
row, the ANETF improves from 11.8 to 22.3 erasures. We can see in
Table~\ref{t1} that, taking codes with the same value $\ell$ of their
layers, their
ANETF and their minimum distance are
roughly correlated when using the parity-check matrix for erasure
decoding. In general, the largest
the minimum distance, the largest the ANETF. An exception is the
2-layer II code with erasure-correcting
capability $(0,0,1,1,1,1,1,2,3,3,3,6)$, which has minimum distance 7
and ANETF 22.7, and the 2-layer EII code with erasure-correcting
capability $(0,0,1,1,1,1,1,1,2,3,4,7)$, which has minimum distance 10
and ANETF 22.6. In spite of the large difference in minimum distance,
the code with minimum distance 7 has slightly better ANETF than the
code with minimum distance 10.

When using the decoding algorithm of Theorem~\ref{theo2},
there is less correlation between minimum distance and
ANETF. For example, the
4-layer II code $(\,((1,1,2),(1,2,3))\,,\,((1,2,3),(1,2,3))\,)$ in
Table~\ref{t1}
has minimum distance 4 and ANETF
15 when using the decoding algorithm of Theorem~\ref{theo2}, while
the 4-layer II code $(\,((0,0,1),(1,1,3))\,,\,((1,1,3),(2,3,6))\,)$
has minimum distance 7 and ANETF
11.8 when using the decoding algorithm. However, we can see that
the ANETF is 17 in the first case and 22.3 in
the second one when decoding using the parity-check matrices.

The first row in Table~\ref{t1} incorporates the (unique) 1-layer
code of rate 62/84 and minimum field size of characteristic~2.
According to Definition~\ref{defmlayer}, it is an $[84,62,23]$ MDS
code over $GF(128)$. If we take the 4-layer code with
erasure-correcting capability
$((\,(0,0,1),(1,1,3))\,,\,((1,1,3),(2,3,6))\,)$ in Table~\ref{t1},
its minimum distance is 7, very far from the MDS bound of 23 (the
MDS bound of course coincides with the ANETF upper bound).
However, the MDS code has no locality (all 84 symbols need to be
accessed in the event of a single erasure) and it requires a field of
size at least the length of the code, which in this case is
$GF(128)$. The ANETF of the 4-layer code, though, is 22.3, which is
at 97\% of the upper bound. In addition, the 4-layer code has multiple
localities in the event of erasures and it is defined over the field
$GF(8)$, much smaller than the field $GF(128)$ required by the MDS code.

Let us point out that computing the ANETF is
related to birthday surprise types of problems~\cite{bgm,gm,gm2,kn}
and obtaining exact
formulae is possible, but in our case they would be too
complicated. Simulations provide good approximations though.

\begin{table}
\begin{center}
\begin{tabular}{|c|c|c|c|c|}
\hline
Code &Erasure-Correcting &Minimum &ANETF&Finite\\
Layer&Capability&Distance & &Field\\
\hline\hline
1&(22)&23&23 ($23^*$)&$GF(128)$\\
\hline\hline
2&(1,1,1,1,1,2,2,2,2,3,3,3)&4&16.6 ($18.6^*$)&$GF(16)$\\
\hline
3&((1,1,2),(1,2,3),(1,2,3),(1,2,3))&4&15.0 ($17.0^*$)&$GF(8)$\\
\hline
4&(((1,1,2),(1,2,3)),((1,2,3),(1,2,3)))&4&15.0 ($17.0^*$)&$GF(8)$\\
\hline\hline
2&(1,1,1,1,1,1,2,2,2,3,3,4)&5&18.8 ($20.8^*$)&$GF(16)$\\
\hline
3&((1,1,2),((1,1,2),(1,2,3)),(1,3,4))&5&16.4 ($20.3^*$)&$GF(8)$\\
\hline
4&(((1,1,2),(1,2,3)),((1,1,2),(1,3,4)))&5&15.4 ($19.6^*$)&$GF(8)$\\
\hline\hline
2&(1,1,1,1,1,1,2,2,2,2,3,5)&6&18.0 ($21.1^*$)&$GF(16)$\\
\hline
3&((1,1,2),(1,1,2),(1,2,3),(1,2,5))&6&16.3 ($20.5^*$)&$GF(8)$\\
\hline
4&(((1,1,2),(1,2,3)),((1,1,2),(1,2,5)))&6&15.4 ($19.9^*$)&$GF(8)$\\
\hline\hline
3&((1,1,2),(1,1,2),(1,2,2),(1,3,5))&6&15.0 ($20.5^*$)&$GF(8)$\\
\hline
4&(((1,1,2),(1,2,2)),((1,1,2),(1,3,5)))&6&14.6 ($20.3^*$)&$GF(8)$\\
\hline\hline
2&(0,0,1,1,1,1,1,2,3,3,3,6)&7&17.5 ($22.7^*$)&$GF(16)$\\
\hline
3&((0,0,1),(1,1,3)),((1,1,3),(2,3,6)))&7&12.4 ($22.4^*$)&$GF(8)$\\
\hline
4&(((0,0,1),(1,1,3)),((1,1,3),(2,3,6)))&7&11.8 ($22.3^*$)&$GF(8)$\\
\hline\hline
2&(0,0,1,1,1,1,1,1,2,3,4,7)&10&15.9 ($22.6^*$)&$GF(16)$\\
\hline
\end{tabular}
\end{center}
\caption{\label{t1} Parameters of some $\ell$-layer II and EII codes of rate 62/84}
\end{table}

\section{Conclusions and Future Work}
\label{conclusions}
We have presented a new definition of Extended Integrated Interleaved
(EII) codes that introduces a slight difference with respect to traditional
definitions in literature. Mainly, we do not require that the nested
codes in the definition have decreasing minimum distances. This
slight difference, though, allows for the construction of
$\ell$-layer EII codes, a new family of codes that establishes a
hierarchy of localities. We showed the properties of the new codes,
in particular, their erasure-correcting capability, dimension,
minimum distance and parity-check matrices. We introduced a new
parameter, the Average Number of Erasures to Failure (ANETF). An upper
bound to the ANETF is the MDS bound. We provided some examples of
constructions approaching the ANETF upper bound, although the codes
are defined over fields much smaller than MDS codes with the same
parameters, have different layers of locality and sparse parity-check
matrices.

Future research will include adapting the constructions of
$\ell$-layer EII codes to codes over any field, as done in~\cite{b}
for 2-layer EII codes, and for decoding of errors as well.

An intriguing topic of research would be to check the performance of $\ell$-layer
EII codes as LDPC codes. In effect, given several $\ell$-layer codes
with the same parameters, the larger $\ell$ is, the sparser the parity-check matrix
of the code is with respect to the other EII codes with lower layer.
For example, take the three next to last
rows in Table~\ref{t1}, corresponding to a 2, 3 and 4-layer code
respectively. For the 2-layer code, all the entries in the
parity-check matrix are non-zero, for the 3-layer code, 86\% of the
entries in the parity-check matrix are non-zero, while for the
4-layer code, 56\% of the entries are non-zero. We can see that as
the layer goes up, the density of non-zero entries goes down
significantly. Of course, more than half of the entries of the
parity-check being
non-zero does not qualify for a code being low density. However,
these are toy examples. Normally LDPC codes involve very long codes.
For example, if we take the 4-layer code whose parity-check matrix is
the $81\times 420$ matrix given by~(\ref{H4}), we can verify that the
density of non-zero entries is only 8.6\%.

\section*{Acknowledgment}
I am very grateful to Wenjie Li, who
provided the inspiration behind many of the ideas in this paper
through our numerous E-mail exchanges.

\end{document}